\documentclass[fleqn,10pt]{wlscirep}

\usepackage{graphicx}
\usepackage{dcolumn}
\usepackage{bm}
\usepackage[titletoc,toc,title]{appendix}
\usepackage{booktabs}
\usepackage{caption}
\usepackage{float}

\usepackage[super,square]{natbib}
\usepackage{color}

\begin{document}

\title{Cell reprogramming modelled as transitions in a hierarchy of cell cycles}             
             
\author[1,*]{Ryan Hannam}
\author[1,2]{Alessia Annibale}
\author[1]{Reimer K\"{u}hn}
\affil[1]{Department of Mathematics, Kings College London, The Strand, London WC2R 2LS,
UK}
\affil[2]{Institute for Mathematical and Molecular Biomedicine, Kings College London, Hodgkin Building, London SE1 1UL,
UK}
\affil[*]{ryan.hannam@kcl.ac.uk}

\begin{abstract}
We construct a model of cell reprogramming (the conversion of fully differentiated cells to a state of pluripotency, known as induced pluripotent stem cells, or iPSCs) which builds on key elements of cell biology viz. cell cycles and cell lineages. Although reprogramming has been demonstrated experimentally, much of the underlying processes governing cell fate decisions remain unknown. This work aims to bridge this gap by modelling cell types as a set of hierarchically related dynamical attractors representing cell cycles. Stages of the cell cycle are characterised by the configuration of gene expression levels, and reprogramming corresponds to triggering transitions between such configurations. Two mechanisms were found for reprogramming in a two level hierarchy: cycle specific perturbations and a noise induced switching. The former corresponds to a \emph{directed} perturbation that induces a transition into a cycle-state of a different cell type in the potency hierarchy (mainly a stem cell) whilst the latter is a priori undirected and could be induced, e.g., by a (stochastic) change in the cellular environment. These reprogramming protocols were found to be effective in large regimes of the parameter space and make specific predictions concerning reprogramming dynamics which are broadly in line with experimental findings.
\end{abstract}

\keywords{Cell reprogramming, cell fate decisions, cell cycle, potency hierarchy, cell lineage, iPSCs.}

\maketitle
\section{\label{sec:Intro}Introduction}
The retrieval of pluripotent cells was first pioneered by John Gurdon in the 1960s, using nuclear transfer to clone a frog using the nuclei of somatic cells extracted from a xenopus tadpol\cite{Gurdon1962}. More recently, cell reprogramming has shown that it is possible to obtain induced pluripotent stem cells (iPSCs), which strongly resemble embryonic stem cells (ES), from somatic cells via the introduction of just 4 transcription factors (Oct3/4, Sox2, Klf4 and c-Myc - now known as the Yamanaka or OSKM factors)\cite{Takahashi2006a, Takahashi2007a}. It has also been demonstrated that nearly all somatic cells can be reprogrammed in this manner\cite{Hanna2009}, suggesting that the ``code" for pluripontency lies in the genome common to all cells of an organism. Once reprogrammed it is possible to guide iPSCs to differentiate into a desired cell type using specific culture conditions \cite{Vierbuchen2012}. Due to their ability to self renew and differentiate into many different cell types, stem cells (including iPSCs) hold great potential for both personalised and regenerative medicine \cite{Cherry2012}. Furthermore, iPSCs can act as a model environment for studying disease and testing drug delivery mechanisms\cite{Cherry2013,Kanherkar2014}. Since the original reprogramming experiments, multiple protocols have been uncovered by replacing certain Yamanaka factors with other proteins or small molecules. For an extensive biomedical review of iPSCs, and how they differ from other stem cells, see Takahashi 2015\cite{Takahashi2015}.

However, despite the great potential of iPSCs, and the evolution of cell reprogramming experiments, much is still unknown about the decisions governing the fate of a cell. Cell fate decisions were first modelled by Waddington using his idea of an epigenetic landscape. This model describes development using the analogy of a ball rolling down a hill from states of high potency to fully differentiated ones. Different cell types are represented as valleys in the landscape, and a cell's fate is determined by the valley which the ball falls into\cite{Gilbert1991}. The number of valleys increases the further the ball moves down the landscape, representing the increasing diversity of cell types during development. Whilst this model provides an interesting metaphor for differentiation, it lacks some key aspects of cell biology, such as cell cycles. Recent experimental work has suggested that cell types can be considered as high dimensional attractors of a gene regulatory network\cite{Huanga}, paving the way for a dynamical systems approach to cell reprogramming.

Many of the current models describing cell fate decisions focus on specific small gene regulatory networks (GRN) that are believed to govern pluripotency or differentiation, or approach the problem from a cell population perspective. For further details of the current mathematical and computational models of cell reprogramming the reader is directed towards the references Morris et al. 2014\cite{Morris2014} and Herberg and Roeder 2015\cite{Herberg2015} respectively.

In this paper a theoretical model is presented which models cell reprogramming in terms of transitions between attractors of a high dimensional dynamical system, describing the transcriptome of a cell. The attractors of the dynamics represent cell cycles of different cell types, which are related to one another in a hierarchical manner.

The rest of this paper is organised as follows: First the theory behind the model is formulated by appealing to a small set of key observations concerning cell chemistry, before being applied to a specific type of hierarchy. Next, evidence for reprogramming in the model is presented and discussed, with the main findings of the work summarised at the end of the paper. The majority of the mathematical details have been relegated to the appendices with the aim of making this article accessible to readers from various backgrounds.

\section{\label{sec:Theory}Theory}
Cells are the fundamental units of structure and reproduction in most organisms\cite{Mazzarello1999}. They are complex and dense building blocks which contain a rich tapestry of biochemical reactions involving a multitude of chemical species (e.g. proteins, sugars, lipids, etc). Metabolic pathways, such as glycolysis, involve many intermediate steps converting the product of one reaction into the substrate for another. Enzyme reactions, like those involved in gycolysis, can be described generally by a set of differential equations, known as the Michaelis-Menten equations\cite{Johnson2011}. Thus, to fully describe the dynamics of cell chemistry one would need to incorporate the Michaelis-Menten equations for all possible reactions into a theory which would describe reaction and diffusion mechanisms, self organisation, biochemical signalling, etc. One component of such a theory would be the transcription of the $N$ genes of the organism's genome ($\sim 25,000$ for humans), which alone represents a vast state space. For example, even if one assumes binary gene expression levels (i.e. genes are either expressed or not expressed), there are $2^{N}$ possible configurations of gene expression levels. A natural question then arises from the complex chemistry of cellular life: How do so many reactions, of so many species, give rise to a comparably low number of different cell types? For example, the human genome is comprised of approximately $25,000$ genes yet only gives rise to around $300$ different cell types. A plausible realisation of this fact is to suppose that stable cell types emerge as attractors of the full reaction dynamics of the cell.

For the purpose of modelling cell reprogramming we propose to construct a reduced model by appeal to the following line of reasoning. Suppose one was able to integrate out all components of the complete theory other than gene expression levels, the result would be a reduced model which will have the following two features: $(i)$ it will involve interactions between genes; $(ii)$ the interactions will exhibit memory effects. The interaction of genes would result in a feedback mechanism that could explain the existence of stable attractors. In the reduced model, memory would be a result of the interplay between genes and proteins. Transcription factors (TFs) are proteins that regulate the expression of genes (through activation/inhibition). These proteins are translated from RNA, which is transcribed from the genes in the cells nucleus. Thus, the expression level of a gene will depend on the previous expression levels through gene regulation. Furthermore, proteins can regulate the genes which they were synthesised from, other genes, and/or combine with other proteins to form complexes which are transcription factors and hence the expression level of a given gene will depend on the previous expression levels of many (or all) genes. Memory is in fact \emph{required} to create dynamic cell cycle attractors with different durations for each of the phases of the cell cycle, in a model based on gene expression levels only.

Based on these observations, we build a \emph{minimal} model that describes cell types in terms of gene expression levels across their cell cycles. We can make simplifications to the reduced model, that do not change the intuition behind, or nature of, the model but make the mathematics easier to work with. One such simplification is the discretisation of time, which allows one to neglect the effects of memory. To do this we measure time in terms of stages passed through the cell cycle (e.g. $G_{1}$, $S$, $G_{2}$, \ldots). This allows one to ignore the different durations of each cycle phase by concentrating on which phase of the cycle a cell is in. Another assumption is that the gene expression levels are binary variables, $n_{i}$ (with $i=1\ldots N$), i.e. genes can exist in one of only two states: they are either expressed or are not. These states may be represented by the binary values $n_{i}=1$ and $n_{i}=0$ respectively, hence the common terminology Boolean, or ``on/off", genes. Again, it is important to stress that these assumptions make the mathematics of the model much simpler, but can be relaxed if a more comprehensive description of cell cycle regulation is required.

A general model for the dynamics of interacting binary genes would have the following form,
\begin{equation}
\label{eq:general_model}
n_{i}(t+1)=\Theta\left[h_{i}(t) -\theta_{i} - T\xi_{i}(t)\right] \,,
\end{equation}
where $n_{i}$ is the gene expression level of the $i^{\textrm{th}}$ gene, with the effect of the gene interactions encoded in a local field, $h_{i}(t)$ of the form,
\begin{equation}
\label{eq:general_local_fields}
h_{i}(t)=  \sum_{j}J_{ij}n_{j}(t) +\sum_{j,k}J_{ijk}n_{j}(t)n_{k}(t)+ \ldots \,.
\end{equation}
Here $J_{ij}$ is the effect of the interaction between genes $i$ and $j$, and $J_{ijk}$ is likewise the effect of the triplet interactions between the 3 genes $i$, $j$ and $k$, (there could also be higher order interactions which are represented by the \ldots). Any constant contributions to the local field, such as self regulation, can be absorbed into the definition of $\theta_{i}$. The $\xi_{i}$ are random variables with zero mean and a suitably normalised variance, which mimic noise to represent the fundamental stochasticity of reaction events. Popular noise models are Gaussian and thermal noise. We use $T$ to vary the strength of the noise. Anticipating
our later choice of the thermal noise model, we will refer to $T$ as to temperature. The $\Theta\left[ x\right]$ is the heaviside step function: $\Theta\left[ x\right]=1$ for $x\geq0$ and $\Theta\left[ x\right]=0$ otherwise. Thus, (\ref{eq:general_model}) states that a gene will be expressed in the next phase of the cell cycle (i.e. $n_{i}(t+1)=1$) if the combined effect of all interactions and stochastic noise at time $t$ exceeds a gene specific threshold, $\theta_{i}$. At each time step every gene expression level is updated according to this rule, and the state of the system is fully described at any time $t$ by the \emph{instantaneous} configuration $\mathbf{n}(t)= \left(n_{1}(t), \ldots, n_{N}(t)\right)$ of gene expression levels.

The network of effective interacting gene expression levels one should consider is a subset of the entire genome. Only the regulatory genes (i.e. the genes that encode for proteins that are transcription factors or form complexes that are transcription factors) need to be considered. Whilst the expression of other genes may contribute to identifying a given cell fate, their expression is driven by that of the regulatory genes. That is the expression of regulatory genes is independent of that of non-regulatory genes. Thus, the total number of genes $N$ in our model should be thought of as the total number of \emph{regulatory} genes.

\subsection{Minimal model}
\label{sec:min_model}
We restrict ourselves to consider a system involving pair interactions only, and simplify matters further by assuming uniform thresholds, i.e. $\theta_{i}=\theta$ $\forall$ $i$. Thus the dynamics of the minimal model is given by the following simple expression,
\begin{equation}
\label{eq:dynamics}
n_{i}(t+1) = \Theta\left[\sum_{j}J_{ij}n_{j}(t)-\theta - T\xi_{i}(t)\right] \,.
\end{equation}
This expression is reminiscent of the models used in the field of neural networks (NNs) for associative memory, with a post synaptic potential (PSP), $h_{i}=\sum_{j}J_{ij}n_{j}$ and a neuron fires ($n_{i}(t+1)=1$) if the PSP exceeds a given threshold $\theta$. In associative memory, configurations of neuronal activity representing some memories are stored in the synaptic efficacies $J_{ij}$, such that they are attractors of the dynamics. The NN is then said to recall a pattern when the system converges to the corresponding configuration from some initial condition. Such NNs are said to be content addressable because the attractor to which they converge is given by the (content of) the initial state. Associative NNs of this type are robust to input errors and hardware failures (such as disruption of the synaptic efficacies and thresholds). Analogously, in our model, specific configurations of gene expression levels, which represent the different cell types of an organism, are stored in the gene interactions, $J_{ij}$, which therefore govern the dynamics. Using a temporal ordering of the cycle state specific configurations, the attractors become dynamic attractors that represent the cell cycles of each cell type (see section on the two level hierarchy for details). It is also desirable that the gene interaction network is robust to variation or errors in gene expression levels, because despite variation in gene expression levels across human individuals and mutations in the human genome, individuals of the population have the same set of cell types.

Hanna et al. demonstrated that nearly all cells can be reprogrammed\cite{Hanna2009}. This suggests that the ``code" for pluripotency lies in the genome that is common to all cell types of a given organism. The analogy between associative memory and cell reprogramming has been made recently in the context of protein interaction networks\cite{Lang2014}. However, that work differs from that presented here due to the absence of cell cycles and the potency hierarchy. In their work, Lang et al. extend Waddington's developmental landscape metaphor into an epigenetic landscape describing the interactions between proteins. On the other hand, by working with gene interactions it is possible to neglect the specific activation/inhibition nature of transcription factors, which will naturally arise from the interaction between genes. Thus, the authors feel that the interaction between genes, as opposed to proteins, is a more natural approach to model cell fates.

\subsection{Cell cycle similarities, lineages and reprogramming}
\begin{figure}[h!]
\begin{center}
\includegraphics[width=0.45\textwidth]{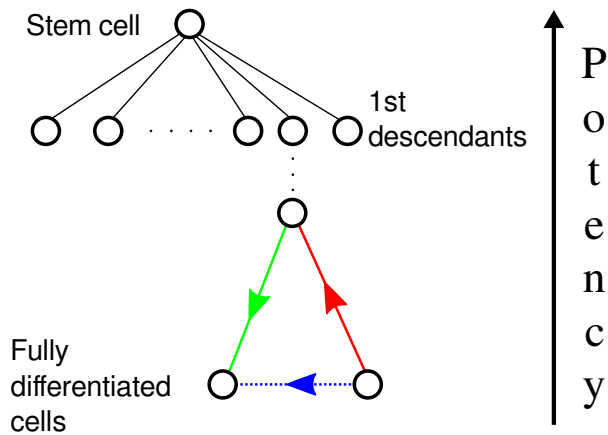}
\end{center}
\caption{\label{fig:cell_hierarchy_schematic}(Colour online.) A schematic diagram of the hierarchy of cell types, in terms of cell potency. Stem cells, e.g. embryonic stem cells (ES) or iPSCs, sit at the top of the hierarchy due to their ability to differentiate into many different cell types. The further down the hierarchy a cell type is, the lower its level of potency (or higher its level of specialisation). Differentiation corresponds to moving down one level of the hierarchy (green arrow); de-differentiation is equivalent to moving up the hierarchy (red arrow); trans-differentiation (blue arrow) corresponds to transitions between cell types of the same level.}
\end{figure}

Cell reprogramming requires transitions between cell fates, which can be either a trans-differentiation or a de-differentiation, i.e, either across or up the potency cascade\cite{Takahashi2015}. With this idea in mind the cell cycles stored in the model are related to one another in a hierarchical manner. Specifically - apart from the stem state which sits at the top of the hierarchy and thus has no ancestor (see figure \ref{fig:cell_hierarchy_schematic}) - the gene expression levels of all cell types are conditionally dependent on their parents. This set up is inspired by the storage of memories, in a Markovian hierarchy, for associative memory NNs\cite{Bos1988}. Parallels between the hierarchical relation of cell cycles and Waddington's epigenetic landscape can be made. However, the present approach does not directly model a landscape of the cell states.

\begin{figure}[h!]
\begin{center}
\includegraphics[width=0.45\textwidth]{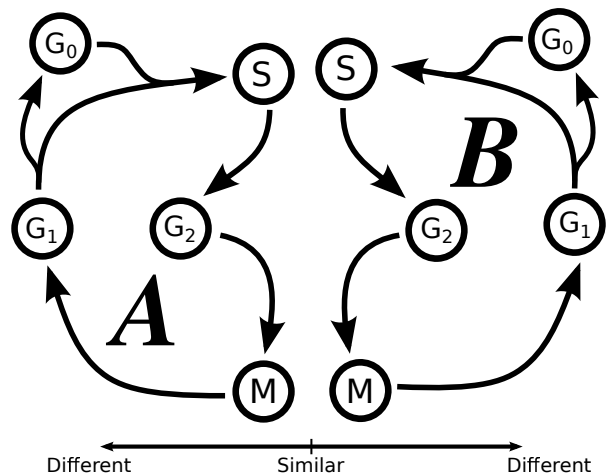}
\end{center}
\caption{\label{fig:cycle_similarity_schematic} (Colour online.) A schematic diagram of the possible similarities between the cell cycles of two different cell types (labelled A and B). The horizontal distances between the analogous phases of the two cell cycles represent the level of similarity - the closer the phases the more similar they are. Two different cell types could be more similar during the S- and/or M-phases, in which the biological processes are broadly similar across different cell types of an organism.}
\end{figure}

It has been demonstrated that protein and mRNA levels vary across the cell cycle\cite{Ly2014}, thus it is reasonable to infer that the gene expression of a cell type also changes throughout its cycle. It is thus plausible to conceive of a situation in which the global expression levels of different cell types are more similar in certain stages of the cell cycle than in others. For example, during the S-phase the gene expression levels could likely be vastly reduced in all cells types (as suggested by Cho 1998\cite{Cho1998}), as the DNA is otherwise occupied through replication. It is also the case that most of an organism's cells undergo the mitotic phase via broadly the same mechanisms. Hence, there could exist (at least) one phase of the cell cycle in which different cell types are more similar than others (see figure \ref{fig:cycle_similarity_schematic}). These stages of the cell cycle would represent a \emph{natural} target in which it is easier to induce switches between cell types, i.e. to reprogramme a cell. This is one of the main hypotheses that we will be testing in the present study.

The authors are aware of only one other model which includes a hierarchy of cell states and the cell cycle. In their model, Artyomov et al. defined a cell type through the expression levels of a small ensemble of master regulatory genes referred to as a \emph{module}\cite{Artyomov2010}. They included the cell cycle as an interplay between the gene expression levels of a cell and the epigenetic state of the cell. On the other hand, we treat each cell type as a dynamic entity, which transitions through different configurations of gene expression levels that correspond to stages of its cell cycle. Each configuration describes the \emph{entire} transcriptome of a cell in a given cycle phase. In this work we do \emph{not} directly model the epigenetics of a cell. However, the similarity between different cell types, during specific phases of the cell cycle discussed above, could be a result of epigenetic changes, such as changes in chromatin structure or the presence of histone markers.

\subsection{\label{sec:2level}Two-level hierarchy}
\begin{figure}[h!]
\begin{center}
\includegraphics[width=0.45\textwidth]{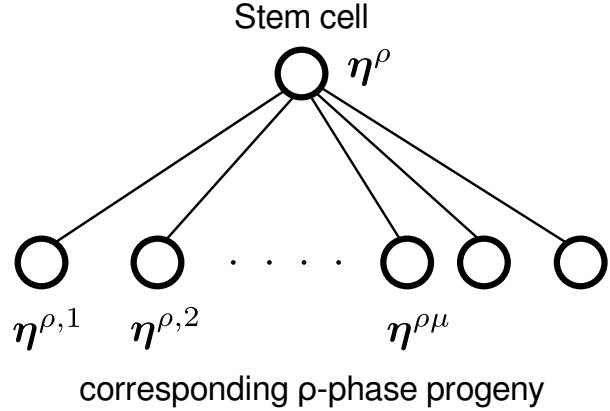}
\end{center}
\caption{\label{fig:2_level_diag} A cell potency hierarchy for a two level system. The stem cell sits at the top of the hierarchy and is given by the configuration of gene expression levels $\boldsymbol\eta^{\rho}$, where the superscript $\rho$ labels the stage of the cell cycle (e.g. S-phase). The second level of the hierarchy consists of $M$ daughter cell configurations. In general the configuration of the daughter cell is given by $\boldsymbol\eta^{\rho\mu}$, where the superscript $\rho$ and $\mu$ label the stage of the cell cycle (e.g. S-phase) and the type of daughter cell respectively (e.g. neuron, B-cell, etc.). A hierarchy of this form exists for every stage of the cell cycle, so that every cell type has the same number of cycle phases.}
\end{figure}

To validate the principles of our approach we apply the neural network model above to a simplified version of the biology. This makes the mathematics easier to implement and keeps the notation simple and transparent. Such simplified scenarios still capture the main principles of the biology, and in practice the mathematics can easily be extended to more realistic systems. We therefore consider a two level hierarchy in which fully differentiated cells are direct descendants, or daughters, of the stem cell (see figure \ref{fig:2_level_diag}). Each of cell cycles are coarse grained into 3 stages, with a single cycle phase made more similar across the different cell types. The coarse graining of the cell cycles was carried out purely for computational efficiency and generalisation to 4 or 5 stage cell cycles is straightforward.

\section{Probabilistic framework}
\label{sec:prob_framework}
\begin{figure*}
 \setlength{\tabcolsep}{0pt}
\begin{tabular}{l r}
\hline
\textbf{Symbol} & \textbf{Biological interpretation}\\ 
$n_{i}(t)$ & Dynamic variable, represents the expression of the $j^{\textrm{th}}$ gene in an arbitrary cell state.\\
$\eta_{i}^{\rho}$ & Expression of $i^{th}$ gene during the $\rho-$phase of the stem cell cycle.\\ 
$\eta_{i}^{\rho\mu}$&  Expression of $i^{th}$ gene during the $\rho$ cycle phase of daughter cell $\mu$.\\
$N$ & Total number of \emph{regulatory} genes in the organism's genome\\
$M$ & Number of daughter cell types.\\
$C$ & Number of phases in the cell cycle.\\
$a^{\rho}$ & Mean stem cell gene expression, or activity, during the $\rho-$phase.\\
$a^{\rho\mu}$ & Mean $\mu-$daughter type gene expression, or activity, during the $\rho-$phase.\\
$a_{\mu}(\eta^{\rho})$ & The average gene expression level in the daughter cell $\mu$, given its value in the stem cell during the same phase, $\rho$.\\
$\gamma^{\rho\mu}$ & Prob. of switching a gene on during differentiation from stem $\rightarrow$ daughter cell $\mu$, during the same phase, $\rho$.\\
$\theta$& Uniform gene activation threshold, the chemical baseline required to activate a gene.\\
$T=\beta^{-1}$ & Effective temperature, a measure of the stochasticity of the dynamics.\\
$\widetilde{m}_{\rho}(\mathbf{n}(t))$ & Dynamical order parameter for the stem cell pattern in the $\rho$ phase.\\
$\widetilde{m}_{\rho\mu}(\mathbf{n}(t))$ & Dynamical order parameter for the $\mu-$daughter cell pattern in the $\rho$ phase.\\
$\mathbf{\widetilde{m}}(t)$& Vector containing all order parameters for both daughter and stem cycles.\\
$m_{\rho}(t)$ & Overlap with the $\rho$ phase of the stem cell cycle.\\
$m_{\rho\mu}(t)$ & Overlap with the $\rho$ phase of the $\mu^{\textrm{th}}$ daughter cell cycle.\\
$J_{ij}$ & The interaction between genes $i$ and $j$.\\
$h_{i}(t)$ & Local field at gene $i$, a measure of influence on gene $i$ by the expression levels of all other genes.\\
$\lambda_{i}(t) $ & Perturbation applied to gene $i$ to reprogramme a cell.\\
$q$ & Probability a perturbation is applied to a gene.\\
$q_{r}$ & The reprogramming threshold.\\
$\xi_{i}(t)$ & Noise variable, a random contribution to the local field, $h_{i}$.\\
\hline
\end{tabular}
 \captionof{table}{\label{table:model_notation}Model notation, along with a brief biological interpretation of the variables used in the model.} 
\end{figure*}

We will now introduce a specific model which implements our generic reasoning within a probabilistic framework. Consider a system of $N$ genes, each labelled by $i=1\ldots N$, and $M$ daughter cell types, labelled by $\mu=1\ldots M$ (for a full list of mathematical notation see table \ref{table:model_notation}). Each cell type, daughter or stem, undergoes a cell cycle of length $C$ and we denote each phase of the cycle by $\rho=1\ldots C$ (with $C+1 \equiv 1$). The expression of the $i^{\textrm{th}}$ gene, in the $\rho^{\textrm{th}}$ phase of the stem cell cycle, is denoted by $\eta_{i}^{\rho} \in \lbrace 0,1\rbrace$, where $\eta_{i}^{\rho}=1$ corresponds to that gene being expressed. We denote by $a^{\rho}$ the fraction of genes expressed  during the $\rho^{\textrm{th}}$ cell cycle phase, also referred to as the activity of that cycle phase. Thus, the probability of the gene expression levels in the $\rho^{\textrm{th}}$ phase of the stem cell cycle are given by
\begin{equation}
\label{eq:stem_bernoulli_distrib}
p(\eta^{\rho})=
\begin{cases}
a^{\rho} & \text{ for } \eta^{\rho}=1 \,, \\
1-a^{\rho} & \text{ for } \eta^{\rho}=0 \,.
\end{cases}
\end{equation}
Then the configuration of the stem cell state, in the $\rho^{\textrm{th}}$ cycle phase, is then given by $\boldsymbol\eta^{\rho}=\left( \eta_{1}^{\rho}, \ldots, \eta_{N}^{\rho}\right)$. Furthermore, for every state, $\boldsymbol\eta^{\rho}$, of the stem cell cycle there is a corresponding set of descendants, in the same stage of the cell cycle, $\boldsymbol\eta^{\rho\mu} = \left( \eta_{1}^{\rho\mu},\ldots,\eta_{N}^{\rho\mu}\right)$. Similar to the stem cell cycle, each daughter cell $\mu$ has an activity given by $a^{\rho\mu}$, which governs the probability of expressing a gene in each stage of the cell cycle
\begin{equation}
\label{eq:daughter_bernoulli_distrib}
p(\eta^{\rho\mu})=
\begin{cases}
a^{\rho\mu} & \text{ for } \eta^{\rho\mu}=1 \,, \\
1-a^{\rho\mu} & \text{ for } \eta^{\rho\mu}=0 \,.
\end{cases}
\end{equation}

We assume the gene expression levels in the stem cell, $\eta_{i}^{\rho}$, are independent, identically distributed (i.i.d) random variables, which implies that the configurations $\boldsymbol\eta^{\rho}$ are independent along the cell cycle. This assumption was made to simplify the mathematics, but may be relaxed if a more comprehensive model is desired. The configurations of the daughter cells, on the other hand, are derived from the corresponding phases of stem cell. We define the probability of turning a gene off during the differentiation-transition from a stem cell to a daughter cell, in the same phases of the cell cycle, with equal activities $a^{\rho}=a^{\rho\mu}$, as $\gamma^{\rho\mu}$. Thus, the transition  probability of a gene being expressed in both the stem and daughter cell (of the same cell cycle phase) is given by $(1-\gamma^{\rho\mu})\frac{a^{\rho\mu}}{a^{\rho}}$, i.e. the ratio of probabilities the gene is on in both states multiplied by the probability it was not turned on in differentiation. The full transition matrix used to construct the gene expression levels of the daughter cells from the same phase of the stem cell cycle can be found in appendix \ref{appendix:W_matrix}. Due to this construction of the daughter cell cycles in terms of the stem cell cycle, the different daughter cell configurations are \emph{conditionally dependent} on the stem cell state. 

The interactions between the gene expression levels of the system should be chosen such that the cell cycles of daughter and stem cells are attractors of the dynamics. To construct the interactions of the model we combine a set of known results from the field of NNs. Hopfield originally showed that multiple configurations can be stored in the synaptic couplings of a neural network using the Hebb rule\cite{Hopfield1982}. It is known that dynamic attractors can be stored in the couplings by adapting the Hebb rule to include a temporal order to the stored configurations, i.e the interactions have a contribution from the current pattern and its successor\cite{Sompolinsky1986,Gutfreund1988}. Thus, in our notation of gene expression levels, a sequence of stem cell cycle phases may be stored in the interactions in the following manner.
\begin{equation}
\label{eq:Jij_cycles}
J_{ij}^{(cycle)}=\dfrac{1}{N}\sum_{\rho=1}^{C}\dfrac{(\eta_{i}^{\rho+1}-a^{\rho+1})(\eta_{j}^{\rho}-a^{\rho})}{a^{\rho}(1-a^{\rho})} \,.
\end{equation}
This choice of interaction ensures that if the gene expression levels evolve according to (\ref{eq:dynamics}) and are initialised, at time $t$, in the configuration $\mathbf{n}(t)=\boldsymbol\eta^{\rho}$, their configuration in the next time step will be $\mathbf{n}(t+1)=\boldsymbol\eta^{\rho+1}$ for sufficiently low noise levels. To ensure the sequence of configurations retrieved by the system is a closed cycle, the successor of the final configuration must be equivalent to the initial configuration. For low activity configurations it is required that one removes the bias from each of the cycle phases, in order to achieve stable limit cycle attractors in the dynamics. Here this is done by subtracting the average gene expression of each of the cell cycle phases, resulting in the contributions from each cycle phase having zero mean.

Information can also be stored in the interactions in a hierarchical manner\cite{Cortes1987,Bos1988,Cortes1988} (equivalent to the structure shown in figure \ref{fig:2_level_diag}). This is done by including contributions from each state in the hierarchy in the interactions. Each pattern must then be weighted by a factor determined by its position in the hierarchy\cite{Parga1986}. Combining these two ingredients, the interactions that stabilise a hierarchy of cell cycles can be written as follows,
\begin{equation}
\label{eq:Jij}
J_{ij}=\dfrac{1}{N}\sum_{\rho=1}^{C}\Biggr\lbrace \dfrac{(\eta_{i}^{\rho+1}-a^{\rho+1})(\eta_{j}^{\rho}-a^{\rho})}{a^{\rho}(1-a^{\rho})} + \sum_{\mu=1}^{M}\dfrac{(\eta_{i}^{\rho+1,\mu}-a_{\mu}(\eta_{i}^{\rho+1}))(\eta_{j}^{\rho\mu}-a_{\mu}(\eta_{j}^{\rho}))}{a^{\rho\mu}(1-a^{\rho\mu})}\Biggr\rbrace \,.
\end{equation}
Here the summations are over cycle phases, $\rho$, and daughter cell types, $\mu$. We chose to remove the bias from the daughter cell type by subtracting the conditional average of the gene expression, $a_{\mu}(\eta^{\rho})=\mathbb{E}\left[\eta^{\rho\mu}|\eta^{\rho}\right]$, i.e. the average gene expression level of the daughter cell given the expression levels in the same cell cycle phase of the stem cell. However, the bias could also be removed using the activity of the daughter cells, $a^{\rho\mu}$, in place of the conditional averages in (\ref{eq:Jij}). The weights in the denominators are the variances of the gene expression levels in the corresponding cell cycle stage for the stem or daughter cells. Note that, if the $\rho+1$ was replaced with $\rho$ in (\ref{eq:Jij}) then this would be the standard prescription for storing a hierarchy of configurations. Since they are included we have in fact stored a \emph{hierarchy of cell cycles}.

Given the form (\ref{eq:Jij}) of the interactions, one can express the local fields $h_i(t)$, appearing in the dynamics (\ref{eq:dynamics}), concisely in terms of a set of macroscopic dynamical order parameters, namely
\begin{equation}
\label{eq:stem_overlap}
\widetilde{m}_{\rho}(t)=\widetilde{m}_{\rho}(\mathbf{n}(t)) = \dfrac{1}{N}\sum_{i=1}^{N}\dfrac{\eta_{i}^{\rho}-a^{\rho}}{a^{\rho}(1-a^{\rho})}n_{i}(t) \,,
\end{equation}
\begin{equation}
\label{eq:daughter_overlap}
\widetilde{m}_{\rho\mu}(t)=\widetilde{m}_{\rho\mu}(\mathbf{n}(t)) = \dfrac{1}{N}\sum_{i=1}^{N}\dfrac{\eta_{i}^{\rho\mu}-a_{\mu}(\eta_{i}^{\rho})}{a^{\rho\mu}(1-a^{\rho\mu})}n_{i}(t) \,,
\end{equation}
giving,
\begin{equation}
\label{eq:local_fields}
 h_{i}(t)=\sum_{\rho}\Biggr\lbrace(\eta_{i}^{\rho+1} -a^{\rho+1})\widetilde{m}_{\rho}(t) + \sum_{\mu}\left[ \eta_{i}^{\rho+1,\mu}-a_{\mu}(\eta_{i}^{\rho+1})\right]\widetilde{m}_{\rho\mu}(t)  \Biggr\rbrace - \theta \,.
\end{equation}
Here we have now absorbed the threshold $\theta$, appearing in \ref{eq:dynamics}, into the definition of $h_i(t)$. The value of $\theta$ is fixed, such that the stable cell cycle attractors exist at sufficiently low noise levels. Note that $h_i(t) = h_i(\widetilde{\mathbf{m}}(t))$, in which $\widetilde{\mathbf{m}}$ is the vector containing the set of all dynamical order parameters $\lbrace\widetilde{m}_\rho(\mathbf{n}(t))\rbrace$ and $\lbrace\widetilde{m}_{\rho\mu}(\mathbf{n}(t))\rbrace$. As a consequence, and by appeal to the law of large numbers in the limit of a large number $N$ of genes, one can formulate the dynamics of our model in closed form \emph{entirely} in terms of the dynamic order parameters, giving
\begin{equation}
\label{eq:general_stem_eqn_motion}
\widetilde{m}_{\rho}(t+1) =\left\langle \dfrac{\eta^{\rho}
-a^{\rho}}{a^{\rho}(1-a^{\rho})} \mathbb{P}\left[\xi\leq \beta h(t)\right] \right\rangle_{\boldsymbol\eta^{\rho},\boldsymbol\eta^{\rho\mu}} \,,
\end{equation}
\begin{equation}
\label{eq:general_daughter_eqn_motion}
\widetilde{m}_{\rho\mu}(t+1) =\left\langle \dfrac{\eta^{\rho\mu}
-a_{\mu}(\eta^{\rho})}{a^{\rho\mu}(1-a^{\rho\mu})}  \mathbb{P}\left[\xi\leq \beta h(t)\right] \right\rangle_{\boldsymbol\eta^{\rho},\boldsymbol\eta^{\rho\mu}} \,,
\end{equation}
provided that $N \gg M$ in this limit. In (\ref{eq:general_stem_eqn_motion}) and (\ref{eq:general_daughter_eqn_motion}) $\beta=1/T$ is the inverse of the noise strength. The $\mathbb{P}(\xi\leq z)$ in (\ref{eq:general_stem_eqn_motion}) and (\ref{eq:general_daughter_eqn_motion}) is the cumulative distribution function (CDF) for the noise probability, $p(\xi)$ (i.e. the probability that $\xi$ will take a value less than or equal to $z$). Popular choices for the $p(\xi)$ are the Gaussian distribution, and the qualitatively and quantitatively similar, thermal distribution $\mathbb{P}(\xi\leq z)=\frac{1}{2}(1+\tanh{\frac{z}{2}} )$. We will use the latter, for which (\ref{eq:general_stem_eqn_motion}) and (\ref{eq:general_daughter_eqn_motion}) can be written in the following form  (for details of this calculation see the appendices \ref{appendix:eqns_of_motion}),
\begin{equation}
\label{eq:stem_eqn_motion}
\widetilde{m}_{\rho}(t+1)=\dfrac{1}{2}\left\langle \dfrac{\eta^{\rho}
-a^{\rho}}{a^{\rho}(1-a^{\rho})} \tanh\left( \dfrac{\beta h(t)}{2}\right)   \right\rangle_{\boldsymbol\eta^{\rho},\boldsymbol\eta^{\rho\mu}} \,,
\end{equation}
\begin{equation}
\label{eq:daughter_eqn_motion}
\widetilde{m}_{\rho\mu}(t+1)=\dfrac{1}{2}\left\langle \dfrac{\eta^{\rho\mu}
-a_{\mu}(\eta^{\rho})}{a^{\rho\mu}(1-a^{\rho\mu})} \tanh\left( \dfrac{\beta h(t)}{2}\right) \right\rangle_{\boldsymbol\eta^{\rho},\boldsymbol\eta^{\rho\mu}} \,,
\end{equation}
where the angle brackets, $\langle\ldots\rangle_{\boldsymbol\eta^{\rho},\boldsymbol\eta^{\rho\mu}}$, represent the average and conditional averages over all stem and daughter cycle states. These equations of motion are easily solved numerically by forward iteration, starting from  suitable initial conditions.

\section{\label{sec:Results}Results}
\begin{figure}
\begin{center}
\includegraphics[width=0.5\textwidth]{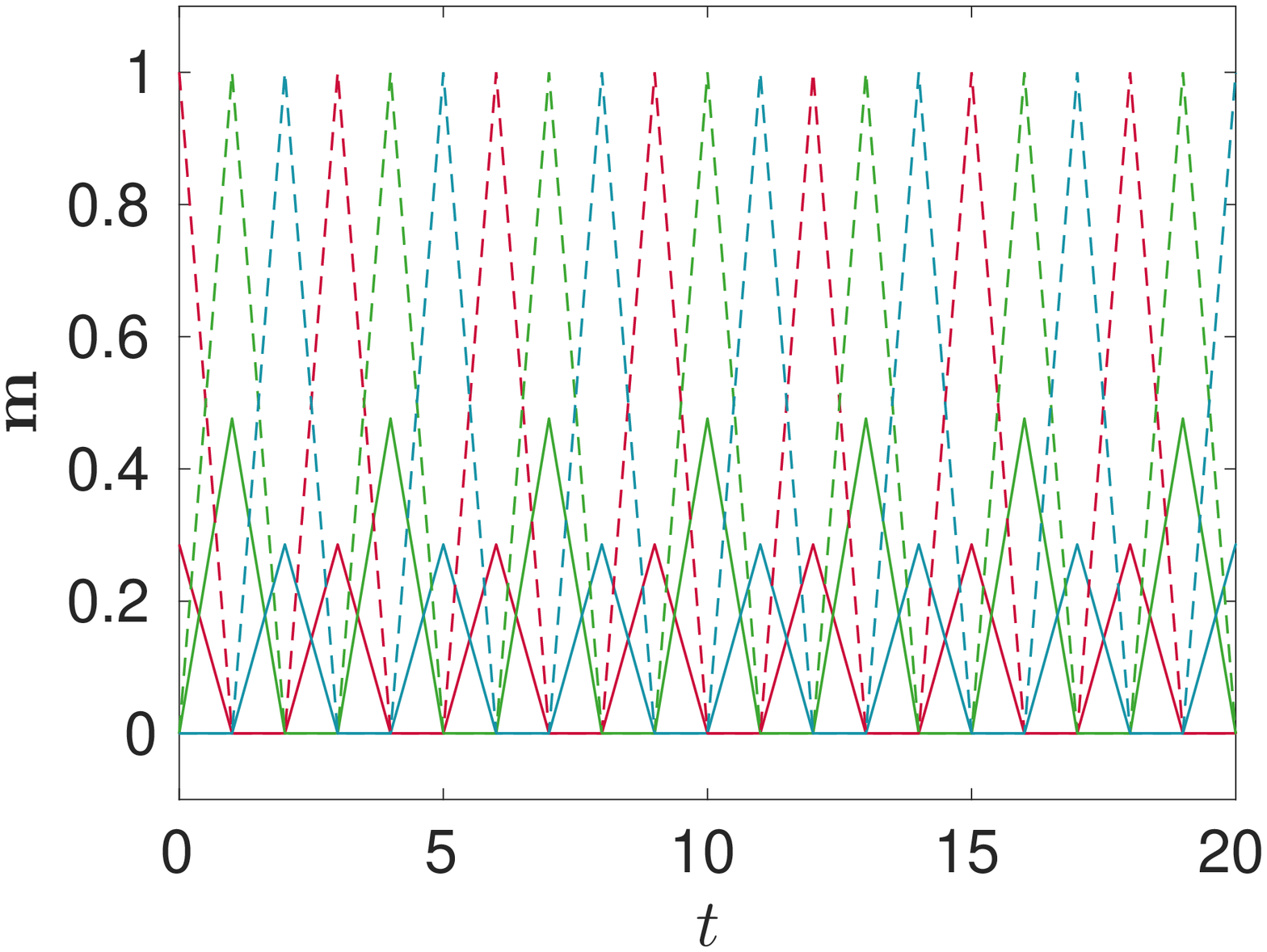}
\end{center}
\caption{\label{fig:Overlap_Trajectories} (Colour online.) Numerical solutions of (\ref{eq:stem_eqn_motion}) and (\ref{eq:daughter_eqn_motion}), at a low effective temperature $T=0.01$, when the system is initialised with a high overlap with the daughter cell cycle. The peaks in $m_{\rho}$ and $m_{\rho\mu}$ correspond to the system passing through states correlated with the different cell cycle stages of the daughter and stem cell - i.e. each peak is a successive $\rho$, and the dashed lines are the overlap with a single daughter type, $\mu$, and the solid lines are the overlap with the stem cell. The following parameters were used: $C=3$, $a^{1}=a^{3}=0.7$, $a^{2}=0.3$, $a^{1\mu}=a^{3\mu}=0.6$, $a^{2\mu}=0.2$, $\gamma^{\rho\mu}=0.2$ for all $\rho$ and $\mu$ and $\theta=0$.}
\end{figure}

In the results that follow a single cell cycle phase was made more similar between the daughter and stem cell cycles. This was done by having a lower value of $a^{\rho}$ and $a^{\rho\mu}$ in that particular phase. The similarity between the cycle phases of the stem and daughter cells can be seen from the covariance between their expression levels,
\begin{equation}
\label{eq:covariance}
\textrm{cov} \left[ \eta^{\rho},\eta^{\rho\mu}\right] = a^{\rho\mu}(1-\gamma^{\rho\mu} - a^{\rho}) \,.
\end{equation}
Thus the probabilities of expressing a gene in both cell cycle stages ($a^{\rho\mu}$ and $a^{\rho}$ respectively) and the probability that a gene is not turned on during differentiation (i.e. $1-\gamma^{\rho\mu}$) govern the similarity between the gene expression levels of the same cycle phase across different cell types. Thus, it is possible to make certain stages of the cell cycle more similar across the two levels of the hierarchy by tuning the parameter values used in (\ref{eq:covariance}). At this point it should also be noted that there are restrictions on the values that $\gamma^{\rho\mu}$ can take, in order for the transition probability from $\boldsymbol\eta^{\rho}$ to $\boldsymbol\eta^{\rho\mu}$ to be correctly defined as a probability (for details see the appendix \ref{appendix:W_matrix}).

Our choice of the parameters is based on the analysis carried out in Ramskold 2009 which suggests that 60-70\% of all genes are expressed in human cells\cite{RamskoLd2009}. In addition, results in the literature suggest that the activity in stem cells is higher than its progeny\cite{Chambers2007}, so we will choose $a^{\rho}>a^{\rho\mu}$ for all $\rho$, $\mu$. This leads us to a choose $a^{\rho}=0.7$ and $a^{\rho\mu}=0.6$ in all cells and all phases except for the phase $\rho=2$, that we aim to make more similar between the stem and daughter cells. In this phase, we chose the parameters $a^{\rho}=0.3$ and $a^{\rho\mu}=0.2$, which lead, via (\ref{eq:covariance}), to a higher similarity between the stem and daughter cells. This is also consistent with the expectation that gene expression is lower in the S-phase due to the genome being occupied with other processes, such as DNA synthesis. The value $\gamma^{\rho\mu}=0.2$ was used for all cell cycle phases, $\rho$, and daughter cell types, $\mu$, and the threshold values were set to zero ($\theta=0$). Unless stated otherwise, these parameter values are used in all of the results and analysis.

For the presentation of the results we use the so-called overlaps, which measure the correlation between the state of the system $\mathbf{n}(t)$ and the gene expression patterns that are characteristic of the cell cycle states of the stem and daughter cells, respectively. They are closely related to the dynamic order parameters, and in fact identical for the stem cell cycles, and are defined as,
\begin{equation}
\label{eq:stem_overlap_rescaled}
m_{\rho}(\mathbf{n}(t))=\widetilde{m}_{\rho}(\mathbf{n}(t)) \,,
\end{equation}
\begin{equation}
\label{eq:daught_overlap_rescaled}
m_{\rho\mu}(\mathbf{n}(t)) = \dfrac{1}{N}\sum_{i=1}^{N}\dfrac{\eta_{i}^{\rho\mu}-a^{\rho\mu}}{a^{\rho\mu}(1-a^{\rho\mu})}n_{i}(t) \,,
\end{equation}
The overlaps are normalized to have values in the interval $\left[-1,1\right]$. An overlap of $m_{\rho\mu}=1$ (or $-1$) means the system is fully correlated (or anti-correlated) with the cell type $\mu$, in the cell cycle phase $\rho$, whereas $m_{\rho\mu}=0$ implies that they are completely uncorrelated. The same is true for the stem cell cycle and corresponding values of $m_{\rho}$.  

In figure \ref{fig:Overlap_Trajectories} we plot the numerical solutions of (\ref{eq:stem_eqn_motion}) and (\ref{eq:daughter_eqn_motion}) when the system is initialised in a daughter cell cycle, at a low noise level, $T=10^{-2}$. Peaks of the dashed line correspond to the system transitioning through the states with a high overlap with the daughter cell cycle phases $m_{\rho\mu}$. Similarly peaks in the solid line correspond to the overlaps with different phases of the stem cell cycle, $m_{\rho}$. 

Because of correlations between the patterns of the cycle states of stem and daughter cells, one observes non-zero mutual overlaps between them (for details see appendix \ref{appendix:Overlaps}). Specifically, if the system is in a (perfect) daughter cell state, $n_i = \eta_i^{\rho\mu}$, the overlap with the corresponding stem cell state is
\begin{equation}
\label{eq:non-zero_overlap}
m_{\rho}(\mathbf{n(t)}=\boldsymbol\eta^{\rho\mu}) = \left\langle \dfrac{\eta^{\rho}-a^{\rho} }{a^{\rho}(1-a^{\rho})}\eta^{\rho\mu} \right\rangle = \dfrac{\textrm{cov} \left[ \eta^{\rho},\eta^{\rho\mu}\right]}{\textrm{var} \left[ \eta^{\rho}\right]} \,.
\end{equation}
Conversely if the system is in a stem cell state, $n_i = \eta_i^{\rho}$, the overlap with the daughter cell state $\eta_i^{\rho\mu}$ is
\begin{equation}
m_{\rho\mu}(\mathbf{n(t)}=\boldsymbol\eta^{\rho}) = \left\langle \dfrac{\eta^{\rho\mu}-a^{\rho\mu} }{a^{\rho\mu}(1-a^{\rho\mu})}\eta^{\rho} \right\rangle = \dfrac{\textrm{cov} \left[ \eta^{\rho},\eta^{\rho\mu}\right]}{\textrm{var} \left[ \eta^{\rho\mu}\right]} \,.
\end{equation}
The peaks corresponding to the phase $\rho=2$ in, figure \ref{fig:Overlap_Trajectories}, are higher, than those corresponding to the other two phases, because the gene activity in this phase was chosen to have a higher covariance, hence a higher value of the overlap (\ref{eq:non-zero_overlap}), between the stem and daughter cells. The other two phases have identical gene activities, hence they have identical overlaps. The initial value of $m_{\rho}(\mathbf{n}(0))$ in figure \ref{fig:Overlap_Trajectories} was determined using (\ref{eq:non-zero_overlap}).

\subsection{\label{sec:NoiseSwitching}Noise induced switching}
\begin{figure*}
\includegraphics[width=0.5\textwidth ]{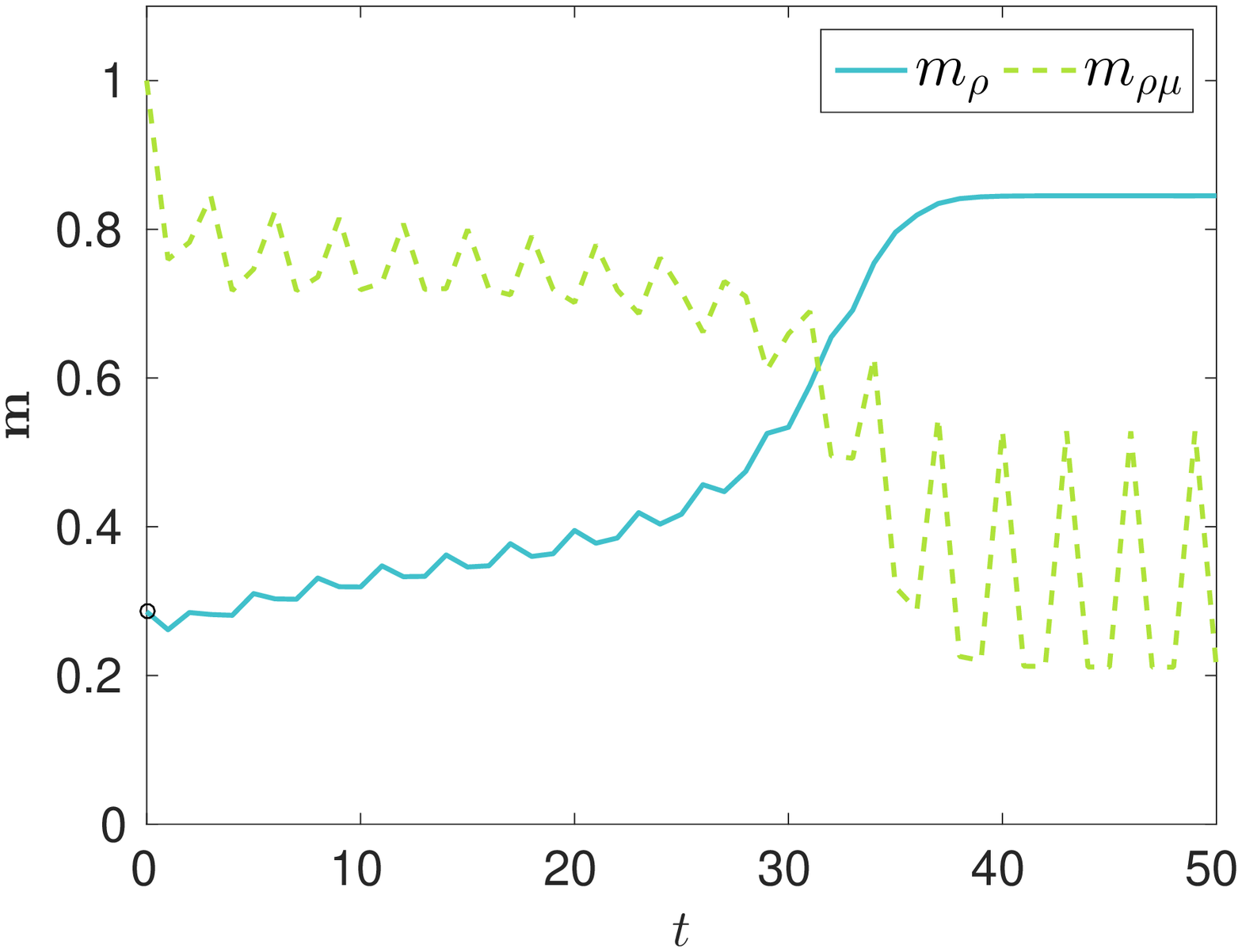}
\includegraphics[width=0.5\textwidth ]{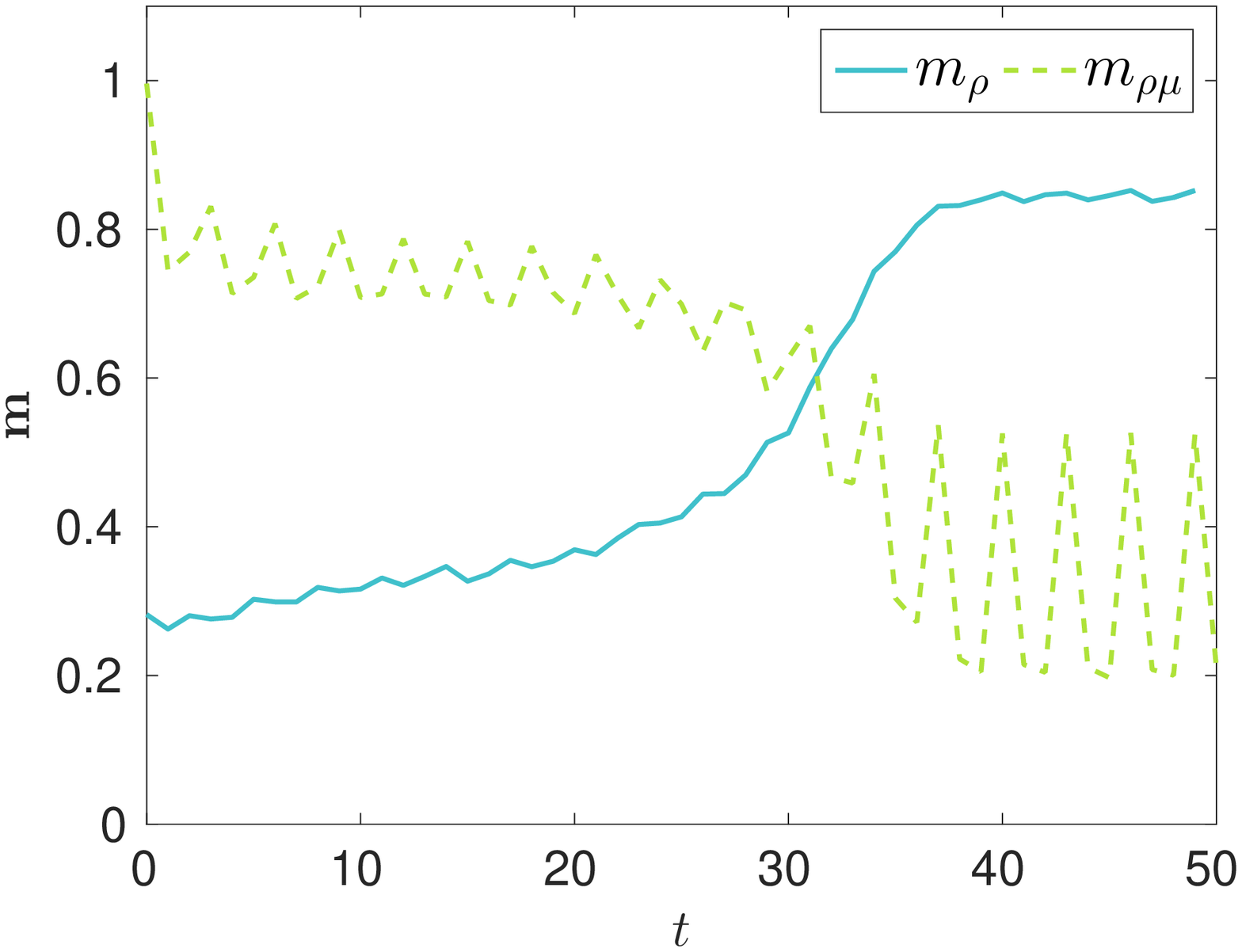}
\caption{\label{fig:Trajectories} (Colour online.) \textbf{Left:} Numerical solutions of (\ref{eq:stem_eqn_motion}) and (\ref{eq:daughter_eqn_motion}), at a noise level $T=0.14$. The initial condition for the overlap with the stem cell was determined using (\ref{eq:non-zero_overlap}). \textbf{Right:} Monte-Carlo simulation dynamics at the same temperature for $N=25,000$ genes. The system was initialised in a configuration with a high overlap with the $\rho=1$ phase of the daughter cell $\mu$, but as the dynamics progress this decays and the system converges to a high value for the overlap with the stem cell cycle. This transition takes multiple generations of the cell cycle and the system passes through an intermediate state with equal overlap with both stem and daughter cell cycles where the two lines intersect. Only the envelope of the trajectories is shown, i.e. the cycle phase, $\rho(t)=1+(t\mod C)$, which the system is expected to be in. For both panels the same parameter values were used as in figure \ref{fig:Overlap_Trajectories}.}
\end{figure*}

At a low noise level, if the system is initialised in a daughter cell it will transition along that cell cycle indefinitely. However, as $T$ is increased above some critical value the noise will take the system away from the daughter cell and it will fall into the attractor corresponding to the stem cell cycle. If the noise is then reduced to a sufficiently low level, the system will become fully correlated with the stem cell cycle. The noise induced transition from the daughter cell cycle to the stem cell cycle is shown in the left panel of figure \ref{fig:Trajectories}, for a value of the temperature $T$ at which the daughter cell cycle is no longer stable. Monte Carlo simulations of the dynamics for $N=25,000$ confirm the validity of our analytic solution formulated in terms of the macroscopic dynamic order parameters in (\ref{eq:stem_eqn_motion}) and (\ref{eq:daughter_eqn_motion}) - right panel of figure \ref{fig:Trajectories}. In this figure we do not plot all time dependent overlaps as in figure \ref{fig:Overlap_Trajectories} but only the ``envelope" of the overlaps defined as the overlaps $m_{\rho(t)}$ and $m_{\rho(t)\mu}$ with the expected cycle state, given by $\rho(t)=1+(t\mod C)$.

The de-differentiation transition takes multiple time steps before a steady state is reached, where the system is in the stem cell cycle attractor. This kind of dynamics is in line with that seen in reprogramming experiments, which take multiple generations of the cell cycle before the iPSCs strongly resemble embryonic stem cells\cite{Takahashi2007a, Hanna2009}.

If, however, the noise level is too high the system quickly loses any correlation with all cell cycles - i.e. all the overlaps become zero. To find the range of noise levels over which it is possible to retrieve the stem cell from the daughter cell cycle one can investigate the stability of the solutions of (\ref{eq:stem_eqn_motion}) and (\ref{eq:daughter_eqn_motion}). Alternatively, we carried out the following numerical experiment: the noise level, $T$, was incremented from zero and at each $T$ the equations of motion were solved numerically, the steady state values of the overlaps ($m_{\rho(t)}$ and $m_{\rho(t)\mu}$) which the dynamics converged to were then recorded. Theses steady state values are plotted against the corresponding noise level in figure \ref{fig:T_diag}. It is clear that above some critical $T$ reprogramming via de-differentiation to the stem cell occurs due to the noise in the system. The value of this critical $T$ will depend on $a^{\rho}$, $a^{\rho\mu}$ and $\gamma^{\rho\mu}$. 

\subsection{\label{sec:DirectPerturbs}Direct perturbations}
\begin{figure}
\begin{center}
\includegraphics[width=0.5\textwidth]{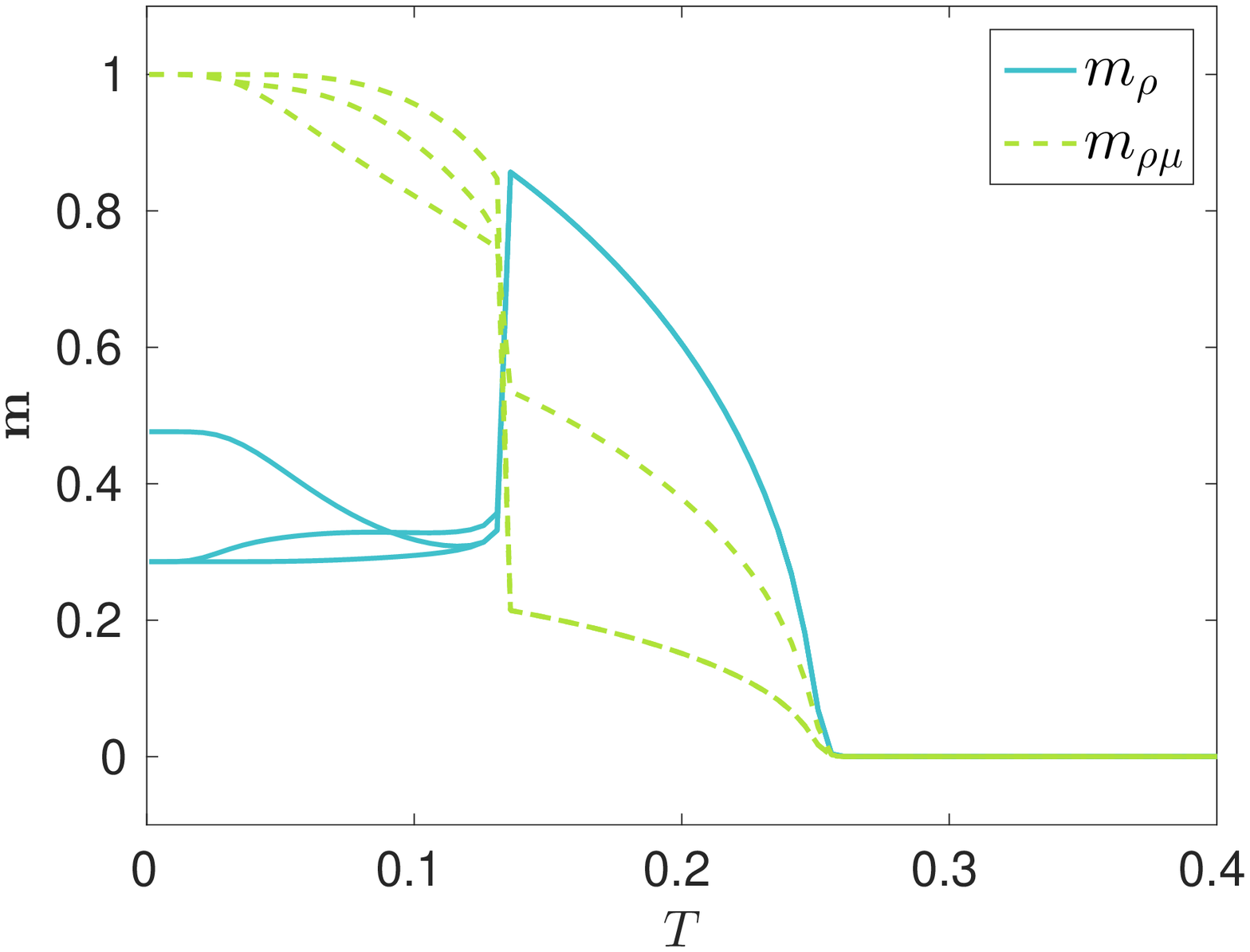}
\end{center}
\caption{\label{fig:T_diag} (Colour online.) Steady state solutions of (\ref{eq:stem_eqn_motion}) and (\ref{eq:daughter_eqn_motion}), showing the overlaps with stem and daughter cell cycle stages ($m_{\rho}$ and $m_{\rho\mu}$ respectively) as a function of noise, $T$. The dashed and solid lines correspond to the daughter and stem cell cycle overlaps respectively. At low $T$ the phases of the stem cell cycle that have the same activity result in identical overlaps $m_{\rho\mu}$. However, as $T$ increases the overlaps for each phase of the daughter and stem cell cycles become distinguishable, before reconverging at critical $T$. Above this critical $T$, the stem cell cycle is retrieved and all $m_{\rho}(\mathbf{n})$ collapse into a single curve, whilst the $\rho=1$ and $3$ curves of $m_{\rho\mu}(\mathbf{n})$ recombine. At sufficiently high values of the overlap, the system becomes uncorrelated with all cell cycles. The same parameter values were used as in figure \ref{fig:Overlap_Trajectories}.}
\end{figure}
The noise induced de-differentiation is different from reprogramming experiments, in which the de-differentiation is due to a direct perturbation using factors that are common to embryonic stem cells (i.e. the Yamanaka factors). Such a directed perturbation can be modelled in our system by introducing an extra contribution 
$\lambda_i(t)$ to the local field, which pushes the system in the direction of the stem cell cycle, and has the form
\begin{equation}
\label{eq:perturbation}
\lambda_{i}(t)= k (\eta_{i}^{\bar{\rho}+1}-a^{\bar{\rho}+1})\widetilde{m}_{\bar{\rho}\mu}(t) c_{i} \,.
\end{equation}
Here $k$ is the strength of the perturbation, $\bar{\rho}$ is the stage of the cycle to which the perturbation is applied, and $c_{i}$ is a logical variable representing whether or not the perturbation is applied to gene $i$ ($c_{i}=1$ with probability $q$, and 0 otherwise).

Since one of the phases of the cell cycle is more similar across different cell types, it is an obvious target for perturbations when attempting to reprogramme a cell. The perturbations should be applied just prior to the most similar phase so as to only minimally disrupt the progression of the cell cycle. So choosing $\bar{\rho}$ as the cycle phase prior to the maximally similar one, is expected to be the optimal reprogramming protocol at a given temperature.

\begin{figure}
\begin{center}
\includegraphics[width=0.5\textwidth]{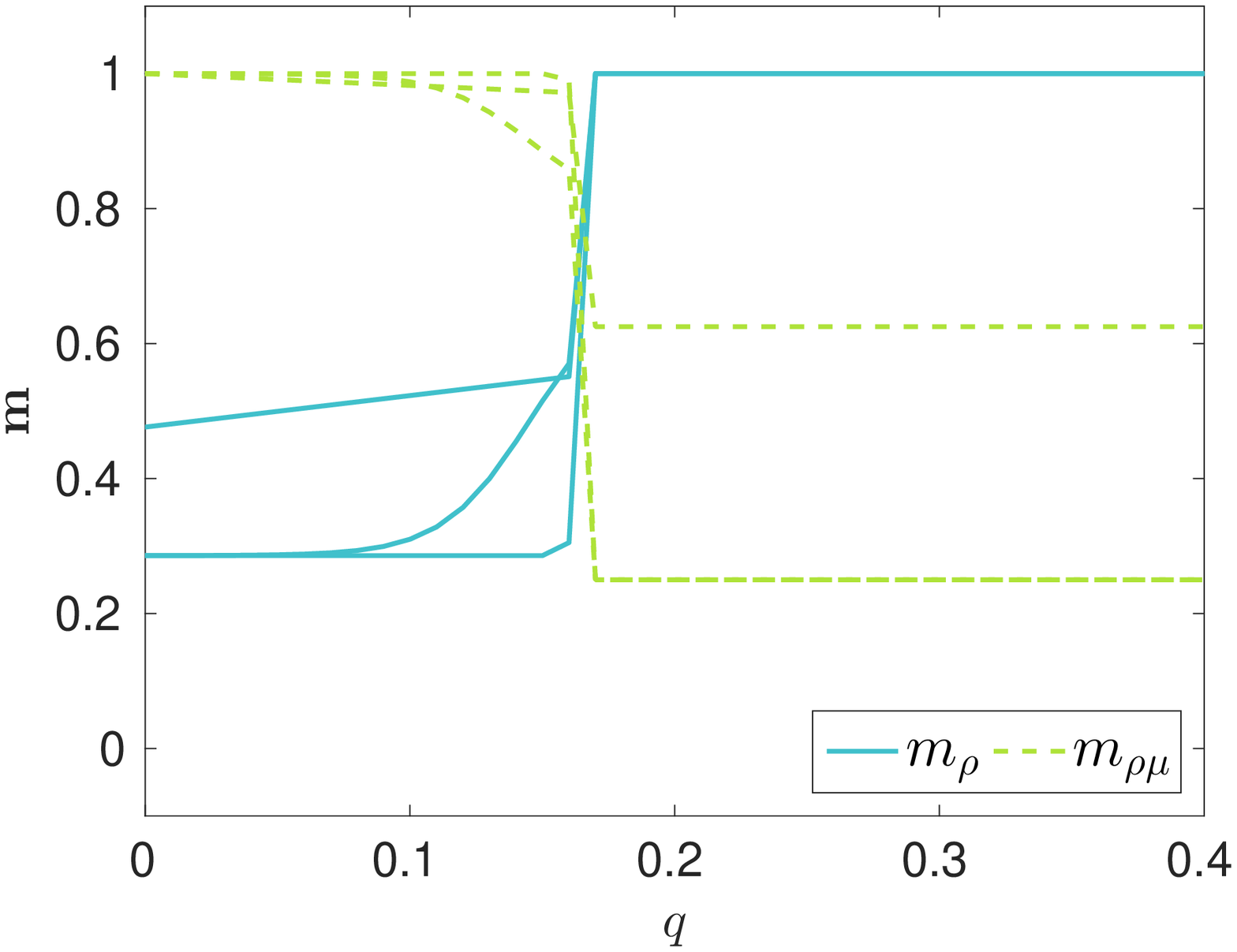}
\end{center}
\caption{\label{fig:q_diag} (Colour online.) Steady state stem and daughter cell cycle overlaps, $m_{\rho}$ and $m_{\rho\mu}$, versus the fraction of genes to which a perturbation of the form (\ref{eq:perturbation}) is applied to, $q$. Here, the perturbations are applied prior to the most similar phase ($\bar{\rho}=1$). The system was kept at a low noise level of $T=0.01$, whilst all other parameter values are the same as in figure \ref{fig:Overlap_Trajectories}. The dashed and solid lines correspond to the daughter and stem cell cycle overlaps respectively. At low $q$ the phases of the stem cell cycle that have the same activity result in identical overlaps $m_{\rho\mu}$. However, as $q$ increases the overlaps for each phase of the daughter and stem cell cycles become distinguishable, before reconverging at critical $q$. Above the critical value of $q$ the stem cell cycle is retrieved and the overlaps for all stem cell phases become identical, whereas overlap with the $\rho=2$ phase of the daughter cell remains separate from the overlaps for the other phases of the daughter cell cycle.}
\end{figure}
In figure \ref{fig:q_diag} we are carrying out the same numerical experiment as in figure \ref{fig:T_diag}, except the probability, $q$, that a perturbation is applied, is incremented rather than the noise level, $T$. This experiment shows that de-differentiation is possible with a directed perturbation even at low noise levels where the daughter cell cycles are stable. The retrieval of stem cell cycle is only possible above some critical value of the fraction of perturbed genes, that we call the reprogramming threshold, $q_{r}$. Because the $\rho=2$ cell cycle stage is more similar across different cell types, perturbations applied to $\bar{\rho}=1$ should have a lower $q_{r}$ value compared with perturbations applied to other phases, i.e. $\bar{\rho}=2$ or $\bar{\rho}=3$. This is indeed borne out by the theory.

\begin{figure*}
\includegraphics[width=0.5\textwidth]{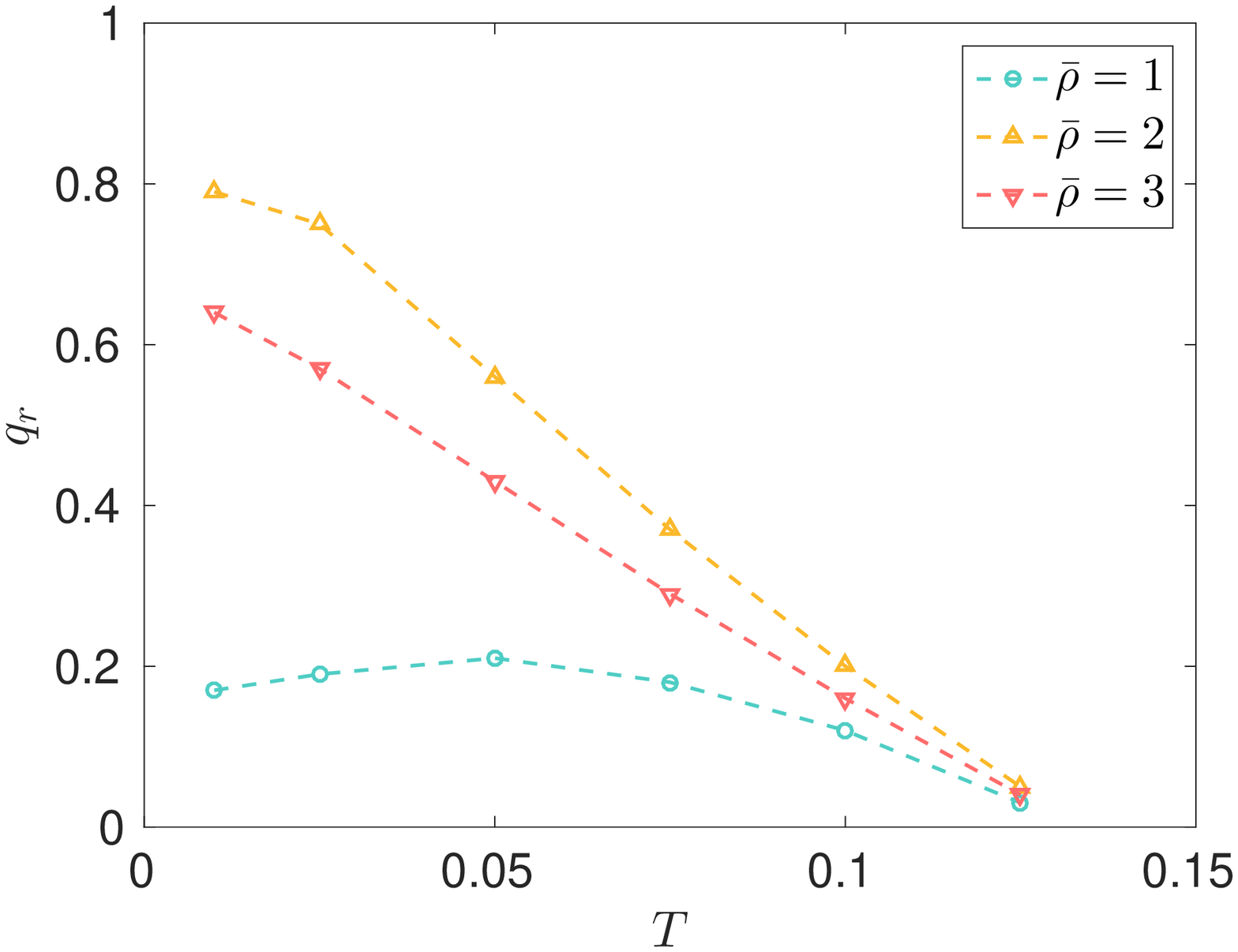}
\includegraphics[width=0.5\textwidth]{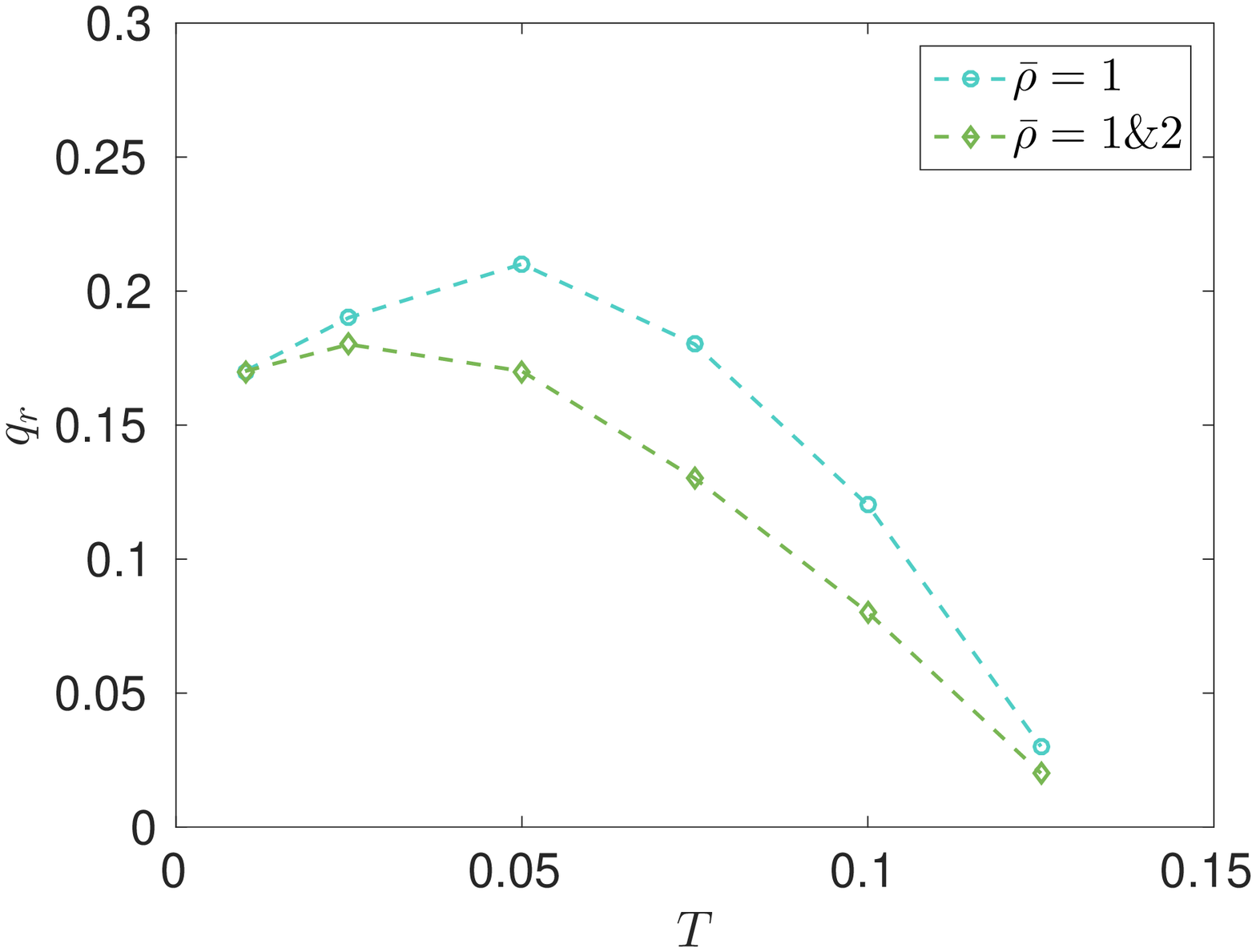}
\caption{\label{fig:T_perturb_dependence} (Colour online.) The fraction, $q_{r}$, of genes that a perturbation of the form \ref{eq:perturbation} is applied to in order to retrieve the stem cell cycle versus noise levels, $T$, increasingly close to that required for the noise induced switching (see figure \ref{fig:T_diag}). The different curves represent different $\bar{\rho}$ protocols for the perturbations. For each protocol a perturbation strength $k=1$ was used, whilst  all other parameter values used are the same as in figure \ref{fig:Overlap_Trajectories}. The relative $q_{r}$ values at a give $T$ can be explained in terms of the Hamming distances between the states involved in the perturbation ($\boldsymbol\eta^{\bar{\rho\mu}}$ and $\boldsymbol\eta^{\bar{\rho}+1}$). As this distance increases so does $q_{r}$. As expected $\bar{\rho}=1$ is the most efficient perturbation for a single target phase - left panel. For the perturbations applied to two phases ($\bar{\rho}=1$ and $2$) - right panel - the perturbations were applied to each phase with an equal probability $q_{r}$.}
\end{figure*}

Increasing the noise level towards the critical $T$ required for noise induced de-differentiation can dramatically change $q_{r}$, see figure \ref{fig:T_perturb_dependence}. The critical value $q_{r}$ has a non-monotonic dependence on the noise level, $T$. This is a direct result of the non-linear nature of the system and the dependence of the perturbation (19) on the dynamical order parameters $\widetilde{m}_{\bar{\rho}\mu}$. 
As expected the $\bar{\rho}=1$ perturbations have the lowest $q_{r}$ values at any given $T$. This is because the $\rho=2$ phase was made to exhibit the largest degree of mutual similarity among cell types, due to a decreased activity in this phase. The fact that the $\bar{\rho}=3$ perturbations have a lower $q_{r}$ than the $\bar{\rho}=2$ perturbations follows from the Hamming distance between the state in which the perturbation is applied and the stem cell state targeted by that perturbation, which is smaller for a perturbation applied in the $3\mu$ state than for a perturbation applied in the 
$2\mu$ state. That is, $d[\bm \eta^{3\mu},\bm \eta^1] < d[\bm \eta^{2\mu},\bm \eta^3]$, where the normalized Hamming distance between states is defined as
\begin{equation}
\label{eq:hamming_distance}
\textrm{d}\left[\boldsymbol\eta^{\bar{\rho}\mu} , \boldsymbol\eta^{\bar{\rho}+1}\right] = \dfrac{1}{N}\sum_{i=1}^{N}\left|\eta_{i}^{\bar{\rho}+1}-\eta_{i}^{\bar{\rho},\mu}\right| \,.
\end{equation}
Hence a higher fraction of genes need to be perturbed to achieve de-differentiation using $\bar{\rho}=2$ compared with $\bar{\rho}=3$. The Hamming distances between different cell cycle states are calculated in terms of the activities in Appendix \ref{appendix:Hamming_distance}.

We have also looked at a case where perturbations of the form (\ref{eq:perturbation}) are acting during multiple stages of the cell cycle in the reprogramming experiments. For example, during the most similar phase and the one prior to it. In this case the effects of each perturbation are combined and $q_{r}$ may decrease compared to applying perturbations to a single phase - right panel of figure \ref{fig:T_perturb_dependence}.

\begin{figure}[h]
\begin{center}
\includegraphics[scale=0.3]{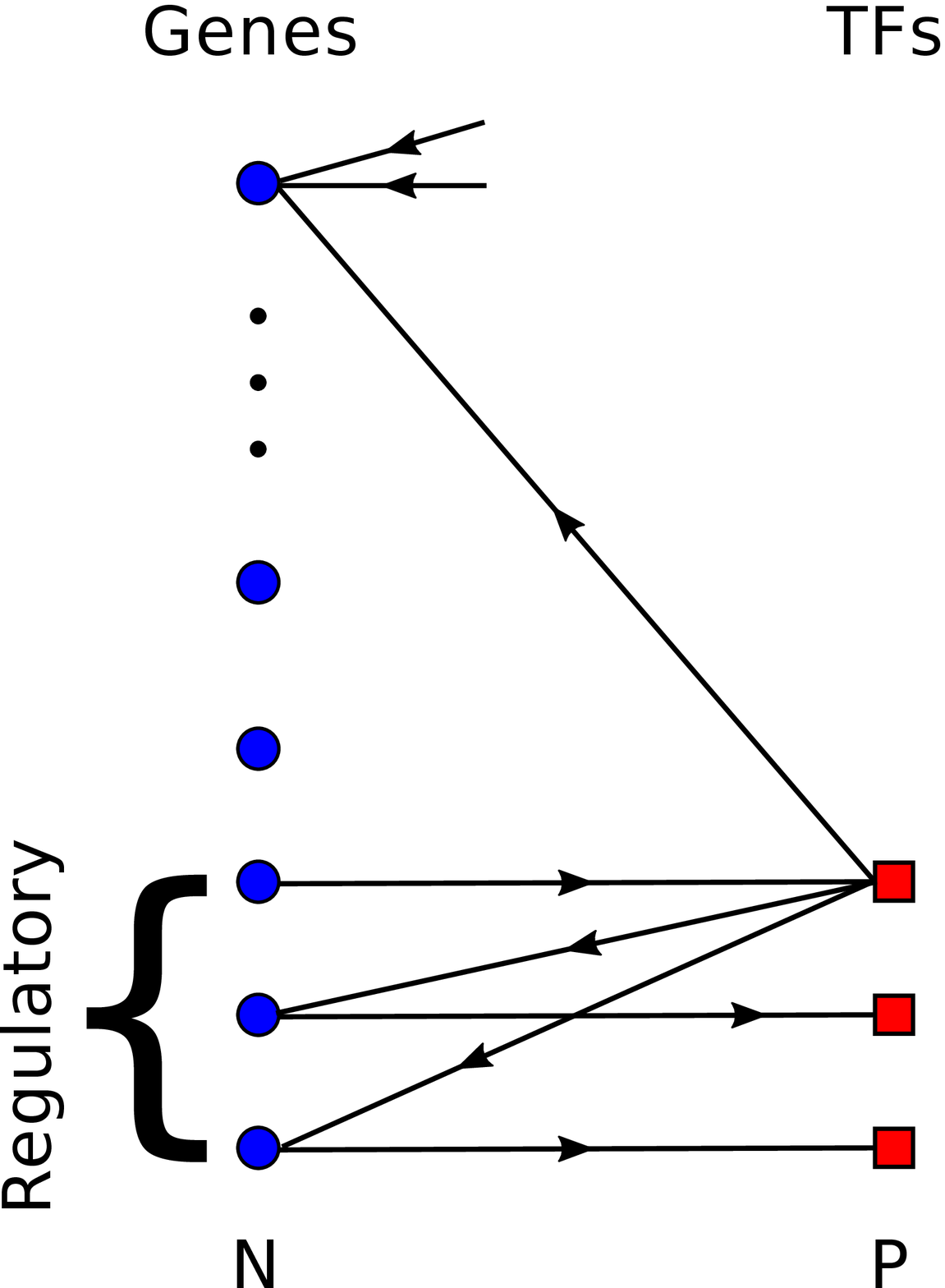}
\end{center}
\caption{\label{fig:bipartite}(Colour online.) A bipartite graph representing the interactions between genes and transcription factors (TFs). The number of TFs scales as $P=\alpha N$ where $\alpha<1$. For conservation, the sum of in-degrees of all genes must equal the sum of out-degrees for all TFs. The number of genes that a TF regulates could then be an order of magnitude larger than the number of TFs that interact with a given gene. Thus introducing a small number of TFs could have a significant effect on the gene expression state of the network. Not all of the nodes and connections of the network are shown.}
\end{figure}

The critical fraction $\mathcal{O}(0.1)$ of genes that need to be perturbed to reprogramme a cell, at low $T$, may initially seem much greater than the four Yamanaka factors introduced in the reprogramming experiments (see figure \ref{fig:q_diag}). However, this order of magnitude is actually in line with the experimental results, as can be seen by considering the following argument. From the $20-25,000$ genes in the human genome only around 10\% are responsible for synthesising transcription factors\cite{Lander2001,Venter2001,Vaquerizas2009}. This imbalance in numbers requires each transcription factor to interact with many more genes than is needed to synthesise it on average. If we consider the interactions between genes and transcription factors as a bipartite graph (see figure \ref{fig:bipartite}), then the TFs have on average an out-degree of $\mathcal{O}(100)$ and the genes have an average in-degree of $\mathcal{O}(10)$ to ensure that there is a conservation in the number of interactions. These numbers agree with the median in- and out-degrees for genes and transcription factors found form a computational analysis of the human gene regulatory network\cite{Narang2015}. Assuming each regulatory gene contributes to the synthesis of a single transcription factor, then perturbing 10\% of the regulatory genes equates to perturbing roughly $250$ genes, which could be achieved by perturbing just 2-3 transcription factors. Thus the fraction $q_{r}=\mathcal{O}(0.1)$ of perturbed gene expression levels is \emph{perfectly in line} with the number of transcription factors used to achieve pluripotency in reprogramming experiments.
~

\section{\label{sec:Summary}Summary}
\label{Summary}
In this paper we presented a general (minimal) model for cell reprogramming as transitions between attractors of a dynamical system. The principles of the model are derived from a set of key facts concerning cell chemistry, which suggest that cell types, and their associated cell cycles, can be considered as attractors of the dynamics of interacting gene expression levels. The specific form of gene interactions used to achieve this goal is inspired by combining two strands of neural network modelling ($i$) storage of (limit) cycles and ($ii$) storage of hierarchically organised attractors.

This paper is intended to provide a proof of concept of this type of modelling approach. We thus decided to investigate the simplest possible hierarchy of cell types that allows us to test our approach, viz. a two-level hierarchy consisting of the stem cell and a single layer of differentiated cells derived from it. Furthermore, we chose to consider only interactions between pairs of binary gene expression levels.

In the present study we show that cell reprogramming is possible using either an undirected approach, which consists of increasing the noise level in the dynamics, or an approach that relies on direct perturbations between specific phases of the cell cycle. Two key non-trivial results appear from our model. Firstly, it takes multiple generations of the cell cycle for a progenitor to be reprogrammed to a stem cell, as it transitions through intermediate states which show similarity with both the initial and final state. Also, a finite fraction of gene expression levels need to be perturbed in order to reprogramme a cell.

We assume that there are states in the cell cycle where mutual similarity in gene expression levels between different cell types is large. These stages of the cell cycle are then natural targets for perturbations to induce changes in cell type. The fraction of genes that need to be perturbed in order to reprogramme a cell depends on the stage of the cell cycle to which the perturbations are applied, as well as the noise level of the system. At low noise levels, this number was in line with that required in the Yamanaka reprogramming experiments and was found to decrease (substantially for $\bar{\rho}=2$ \& $3$) with increasing noise levels. The ``true" noise level of a cell is difficult to quantify, but our model allows for reprogramming in both low and high noise regimes. 

As far as the authors are aware, gene expression levels in different levels of the cell potency hierarchy or in different phases of the cell cycle are still not well characterized. In the present study we have used a scenario where gene expression levels in differentiated
cells are slightly lower than in a stem cell during the same phase of the cell cycle, and we have taken one of the cycle phases to have lower levels of gene expression than the others (thereby increasing mutual similarities of different cells during this cycle phase). We have checked that de-differentiation along the two different routes, noise induced and via directed perturbations, does not depend on these specific choices, although details, such as critical thresholds for reprogramming, do change as scenarios are modified.

At the time of writing this article, the authors have become aware of another study that models the cell cycle as configurations of gene expression patterns using a cyclic Hopfield model\cite{Szedlak2017a}. However, our study goes beyond modelling a single cell cycle and studies transitions between different cell types.

There are some limitations to our modelling approach. Firstly, de-differentiation is modelled in terms of the entire genome of a cell, whereas only a subset of gene expression levels could be responsible for cell fates and differentiation. Our model can be adapted to consider only a sub-network of gene expression levels which are responsible for cell fates. In order to maintain the large diversity of cell types, higher order interactions will be required to stabilise and store a large number of attractors $M$. This is biologically reasonable, since proteins expressed from multiple genes can form complexes, which are transcription factors, and genes can often require proteins binding to promoter sites and enhancer regions before the gene is expressed. Using discrete time dynamics excludes the possibility of variability in gene expression levels in a given cell cycle phase. Therefore any in-cycle dynamics is missed, such as any cell signalling cascades. Finally, we only use rough estimates for the average gene expression levels in numerical experiments and simulations. 

One possibility for extending our model would be to relax the choice of independent cell cycle states. Correlated cell cycle phases can be incorporated into the two level hierarchy by changing the way in which the cell cycles are constructed in the model. One could achieve this using a three level hierarchy to store all cell cycles, whilst maintaining the feature that all descendants are a single differentiation from the stem cell cycle. In this situation the root of the hierarchy would be a template of the stem cell expression levels, the second level would then be constructed from this and represent each stage of the stem cell cycle. The newly included third level would consist of each daughter cell type branching off from the corresponding stem cell cycle phase. Such a set up would then be analogous to a two level hierarchy for each cell cycle phase with correlations in the gene expression levels along the cell cycles of each cell type.

In the future the authors hope to work on applying this model to real data in order to test its validity and aid in the design of reprogramming protocols. However, since there is still much to be learnt about cell reprogramming and the decisions of cell fates in developmental biology, the model has been presented here to encourage the discussion between experimentalists and those from a more theoretical background (such as statistical physics, the field which inspires much of this work), because clearly a mixture of these skills will be required in driving our understanding of cell fates and reprogramming forward.\\
\\
\textbf{Competing interests:} We declare we have no competing interests.\\
\textbf{Funding:} R.H. is supported by the EPSRC Centre for Doctoral Training in Cross-Disciplinary Approaches to Non-Equilibrium Systems (CANES, EP/L015854/1).\\
\textbf{Acknowledgements:} The authors would like to thank Attila Csik{\'a}sz-Nagy and Ignacio Sancho-Martinez for the insightful discussions during the formulation of the model.\\
\textbf{Authors' contributions:} All authors conceived the model and designed the numerical experiments. R.H. performed the numerical experiments and analysed the data. All authors wrote the paper.

\bibliography{manuscript_bibliography}


\appendix
\section{Appendix A: Differentiation transition matrices}
\label{appendix:W_matrix}
To derive the daughter cell expression levels from the stem cells a transition matrix, $\mathbf{W}$, was used. When $\mathbf{W}$ is applied to the probability distribution of the a stem cell phase the result corresponds to the distribution of the daughter state, i.e. $\mathbf{W}\begin{bmatrix}
a^{\rho} \\ 1-a^{\rho}
\end{bmatrix}=\begin{bmatrix}
a^{\rho\mu} \\ 1-a^{\rho\mu}
\end{bmatrix}$. If we define the probability gene is switched on in the differentiation from the stem to daughter cell as $\gamma^{\rho\mu}$, for the same activities in each cell cycle phase ($a^{\rho}=a^{\rho\mu}$), then we can define $\mathbf{W}$ as the following matrix.
\begin{equation}
\label{eq:W_matrix}
 W(\eta^{\rho\mu}|\eta^{\rho})=
 \begin{bmatrix}
 (1-\gamma^{\rho\mu})\dfrac{a^{\rho\mu}}{a^{\rho}} & \dfrac{\gamma^{\rho\mu}a^{\rho\mu}}{1-a^{\rho}}\\
 1-(1-\gamma^{\rho\mu})\dfrac{a^{\rho\mu}}{a^{\rho}}  & 1-\left(\dfrac{\gamma^{\rho\mu}a^{\rho\mu}}{1-a^{\rho}}\right)
 \end{bmatrix} \,.
\end{equation} 
To have (\ref{eq:W_matrix}) defined as a transition matrix, its columns must sum to one and each element  $W(\eta^{\rho\mu}|\eta^{\rho})\in \left[0,1\right]$. Since $a^{\rho}$ and $a^{\rho\mu} \in \left[0,1\right]$ by definition, we must obey the following constraint on $\gamma^{\rho\mu}$ for itself and (\ref{eq:W_matrix}) to be defined as probabilities,
\begin{equation}
\label{eq:gamma_restriction}
\max\left[0,\dfrac{a^{\rho\mu}-a^{\rho}}{a^{\rho\mu}}\right] \leq \gamma^{\rho\mu} \leq \min\left[1, \dfrac{1-a^{\rho}}{a^{\rho\mu}}\right] \,.
\end{equation}

\section{Appendix B: Equations of motion}
\label{appendix:eqns_of_motion}
 In each time step the state of the system is updated based on the field local fields of each site. That is, the expression level of gene $i$, at time $t$, depends on the value of the field at time $t-1$, i.e.
\begin{equation}
\label{eq:dynamical_update}
n_{i}(t+1)=\Theta\left[h_{i}(\mathbf{m}(t))-\xi_{i}(t)\right]\,,
\end{equation}
where $\Theta(x)$ is the Heaviside step function ($\Theta(x)=0$ for $x\leq 0$ and $\Theta(x)=1$ for $x>0$), and $\xi_{i}(t)$ is thermal noise at the site $i$ (with $\mathbb{P}\left[\xi_{i}(t)<z\right] = \frac{1}{2}\left[1+\tanh(\beta z/2)\right]$). 

The expected value of site $i$ can be obtained by averaging (\ref{eq:dynamical_update}),
\begin{align}
\langle n_{i}(t+1)\rangle &= \mathbb{P}\left[\Theta\left[ h_{i}(\mathbf{m}(t))-\xi_{i}(t)\right] >0\right]  \\
&= \mathbb{P}\left[\xi_{i}(t) < h_{i}(\mathbf{m}(t))\right]  \\
&= \dfrac{1}{2}\left[1+\tanh\left(\dfrac{\beta h_{i}(\mathbf{m}(t))}{2}\right)\right]\,.
\end{align}
Then using the definitions of the dynamics order parameters (\ref{eq:stem_overlap}) and (\ref{eq:daughter_overlap}), following expressions can be obtained, when the $N\rightarrow\infty$ limit is taken, by making use of the law of large numbers:
\begin{align}
\widetilde{m}_{\rho}(t+1)=\dfrac{1}{2}\left\langle \dfrac{\eta^{\rho}
-a^{\rho}}{a^{\rho}(1-a^{\rho})} \tanh\left( \dfrac{\beta h(t)}{2}\right)   \right\rangle_{\boldsymbol\eta^{\rho},\boldsymbol\eta^{\rho\mu}} \,, \\
\widetilde{m}_{\rho\mu}(t+1)=\dfrac{1}{2}\left\langle \dfrac{\eta^{\rho\mu}
-a_{\mu}(\eta^{\rho})}{a^{\rho\mu}(1-a^{\rho\mu})} \tanh\left( \dfrac{\beta h(t)}{2}\right) \right\rangle_{\boldsymbol\eta^{\rho},\boldsymbol\eta^{\rho\mu}} \,,
\end{align}
where $\langle\ldots\rangle_{\boldsymbol\eta^{\rho},\boldsymbol\eta^{\rho\mu}}$ is shorthand for an average over the statistics of (correlated) stem and daughter cell expression levels throughout their cycles,
$$ \prod_{\rho=1}^{C}\left[ \sum_{\eta^{\rho}\in\lbrace 0,1\rbrace}p(\eta^{\rho}) \prod_{\mu=1}^{M} \left[ \sum_{\eta^{\rho\mu}\in\lbrace 0,1\rbrace}W(\eta^{\rho\mu}|\eta^{\rho})(\ldots) \right]\right] \,.$$

A more rigorous calculation to determine these equations of motion can be done by following the reasoning in Coolen et al\cite{CoolenBook}. 
\section{Appendix C: Order parameters for specific cell cycle configurations}
\label{appendix:Overlaps}
This appendix contains the calculation of the order parameters, $\widetilde{m}_{\rho}(\mathbf{n}(t))$ and $\widetilde{m}_{\rho\mu}(\mathbf{n}(t))$, when the system is in different generations of the cell hierarchy. First, $\widetilde{m}_{\rho}(\mathbf{n}(t))$ when the system is in the daughter cell cycle configuration. Equation (\ref{eq:stem_overlap}) can be rewritten as follows by making use of the law of large numbers. ($N\rightarrow\infty$) with $n_{i}=\eta_{i}^{\bar{\rho}\mu}$,
\begin{equation}
\label{eq:daughter_overlap_with_stem}
\widetilde{m}_{\rho}(\mathbf{n}=\boldsymbol\eta^{\bar{\rho}\mu})=\left\langle \dfrac{\eta^{\rho}-a^{\rho}}{a^{\rho}(1-a^{\rho})}\eta^{\bar{\rho}\mu}\right\rangle \,.
\end{equation}
If $\bar{\rho}\neq\rho$ then the daughter cell cycle phase is independent of the stem cell cycle phase and the expectation value factorises,
\begin{equation}
\widetilde{m}_{\rho}(\mathbf{n}=\boldsymbol\eta^{\bar{\rho}\mu})=\left\langle \dfrac{\eta^{\rho}-a^{\rho}}{a^{\rho}(1-a^{\rho})}\right\rangle \left\langle\eta^{\bar{\rho}\mu}\right\rangle \,,
\end{equation}
and $\widetilde{m}_{\rho}(\mathbf{n}=\boldsymbol\eta^{\bar{\rho}\mu})=0$ since $\langle\eta^{\rho}-a^{\rho}\rangle=0$.

However, if $\bar{\rho}=\rho$ then (\ref{eq:daughter_overlap_with_stem}) can be written as,
\begin{align}
\nonumber \widetilde{m}_{\rho}(\mathbf{n}=\boldsymbol\eta^{\rho\mu}) &=\mathbb{E}\left[ \mathbb{E}\left[ \dfrac{\eta^{\rho}-a^{\rho}}{a^{\rho}(1-a^{\rho})}\eta^{\rho\mu} \biggr|\eta^{\rho\mu}\right] \right] \\
\nonumber &= \sum_{\eta^{\rho}} \dfrac{\eta^{\rho}-a^{\rho}}{a^{\rho}(1-a^{\rho})}p(\eta^{\rho}) \sum_{\eta^{\rho\mu}} \mathbb{W}(\eta^{\rho\mu}|\eta^{\rho})\eta^{\rho} \\
&= \dfrac{a^{\rho\mu}}{a^{\rho}}\left(\dfrac{1-\gamma^{\rho\mu}-a^{\rho}}{1-a^{\rho}}\right)\,,
\end{align}
where $\mathbb{E}[x]$ and $\mathbb{E}[x|y]$ represents the expectation value of $x$ and of $x$ given $y$, respectively. Thus, provided $a^{\rho\mu}>0$ and $\gamma^{\rho\mu}<1$, there is a non-zero value for the order parameter. This can be rewritten by noticing that the numerator is the covariance between $\eta^{\rho}$ and $\eta^{\rho\mu}$ and the denominator is the variance of $\eta^{\rho}$. Thus, for $n_{i}=\eta_{i}^{\rho\mu}$ $\forall i$,
\begin{equation}
\widetilde{m}_{\rho}(\mathbf{n}=\boldsymbol\eta^{\bar{\rho}\mu})= \dfrac{\textrm{cov} \left[ \eta^{\rho},\eta^{\rho\mu}\right]}{\textrm{var} \left[ \eta^{\rho}\right]}\,.
\end{equation}

Next, $\widetilde{m}_{\rho\mu}(\mathbf{n}(t))$ when the system is in the stem cell cycle configuration. Following similar reasoning to the above, (\ref{eq:daughter_overlap}) can be written as,
\begin{equation}
\label{eq:stem_overlap_with_daughter}
\widetilde{m}_{\rho\mu}(\mathbf{n}=\boldsymbol\eta^{\bar{\rho}})=\left\langle \dfrac{\eta^{\rho\mu}-a_{\mu}(\eta^{\rho})}{a^{\rho\mu}(1-a^{\rho\mu})}\eta^{\bar{\rho}}\right\rangle\,,
\end{equation}
where again it is trivial that $m_{\rho\mu}=0$ due to independence if $\bar{\rho}\neq\rho$.

For the case $\bar{\rho}=\rho$,
\begin{align}
\nonumber \widetilde{m}_{\rho\mu}(\mathbf{n}=\boldsymbol\eta^{\bar{\rho}})&=\mathbb{E}\left[ \mathbb{E}\left[ \dfrac{\eta^{\rho\mu}-a_{\mu}(\eta^{\rho})}{a^{\rho\mu}(1-a^{\rho\mu})}\eta^{\bar{\rho}} \biggr|\eta^{\rho\mu}\right] \right] \\
&=\sum_{\eta^{\bar{\rho}}}\eta^{\bar{\rho}}p(\eta^{\bar{\rho}})\sum_{\eta^{\rho\mu}}\dfrac{(\eta^{\rho\mu}-a_{\mu}(\eta^{\rho}))\mathbb{W}(\eta^{\rho\mu}|\eta^{\rho})}{a^{\rho\mu}(1-a^{\rho\mu})} \,.
\end{align}
Then since $\sum_{\eta^{\rho\mu}}\mathbb{W}(\eta^{\rho\mu}|\eta^{\rho})=1$ and $\sum_{\eta^{\rho\mu}}\mathbb{W}(\eta^{\rho\mu}|\eta^{\rho})\eta^{\rho\mu}=a_{\mu}(\eta^{\rho})$ by their definitions, $\widetilde{m}_{\rho\mu}(\mathbf{n}=\boldsymbol\eta^{\bar{\rho}})=0$ even if $\bar{\rho}=\rho$. Thus, during any phase of the stem cell cycle all of the order parameters for each of the daughter cell cycle phases vanish.

\section{Appendix D: Normalised Hamming distance}
\label{appendix:Hamming_distance}
The normalised Hamming distance between two vectors $\mathbf{x}=(x_{1},x_{2}\ldots x_{N})$ and $\mathbf{y}=(y_{1},y_{2}\ldots y_{N})$ is defined as follows,
\begin{equation}
\textrm{d}\left[\mathbf{x},\mathbf{y}\right] = \dfrac{1}{N}\sum_{i=1}^{N}\left|y_{i}-x_{i} \right| \,.
\end{equation}
In the large $N$ limit we can replace the sum over $N$ using the law of large numbers to obtain, 
\begin{equation}
\textrm{d}\left[\mathbf{x},\mathbf{y}\right] =\left\langle\left|y_{i}-x_{i} \right|\right\rangle.
\end{equation}
Thus, for the same phase of the cell cycle the Hamming distance between the stem and daughter cell cycles is given by,
\begin{equation}
\textrm{d}\left[\boldsymbol\eta^{\rho\mu},\boldsymbol\eta^{\rho}\right] =a^{\rho}-a^{\rho\mu}+2\gamma^{\rho\mu}a^{\rho\mu} \,,
\end{equation}
where the averages were preformed over the conditional the joint probability distribution $p(\eta^{\rho},\eta^{\rho\mu})=\mathbb{W}(\eta^{\rho\mu}|\eta^{\rho})p(\eta^{\rho})$. Similarly the Hamming distance between a daughter cell cycle phase and the next phase of the stem cell cycle is given by,
\begin{equation}
\textrm{d}\left[\boldsymbol\eta^{\rho\mu},\boldsymbol\eta^{\rho+1}\right] =a^{\rho+1}-a^{\rho\mu}-2a^{\rho\mu}a^{\rho+1} \,,
\end{equation}
where the average was preformed over the joint probability, $p(\eta^{\rho+1},\eta^{\rho\mu})=p(\eta^{\rho+1})p(\eta^{\rho\mu})$ where the factorisation is due to the independence between the different cell cycle phases. 
\end{document}